\documentclass{article}

\usepackage{fullpage}
\usepackage[normalem]{ulem}
\usepackage{textalpha}
\usepackage{amsthm, amsfonts, amssymb, extpfeil, bbm,url,microtype,mathtools,mathrsfs,graphicx,float,color}
\usepackage[ruled]{algorithm2e}
\usepackage{algpseudocode}
\usepackage{physics}
\usepackage{lmodern}
\usepackage[hidelinks]{hyperref}
\usepackage{tikz}
\usepackage{subfig}
\usepackage[T1]{fontenc}
\usepackage[USenglish]{babel}
\usepackage{listings}
\usepackage{authblk}
\usepackage{pgfplots}
\usepackage{cleveref}

\usepackage[qm]{qcircuit}

\usetikzlibrary{shapes.geometric,angles, 
patterns,positioning, fit, calc,intersections,quotes}
\pgfplotsset{compat=1.18} 
\usepgfplotslibrary{fillbetween}
 
\newcommand{\tarc}{\mbox{\large$\frown$}}
\newcommand{\arctikz}[2][-3ex]{{#2}{\kern #1{\raisebox{1.5ex}{\tarc}}}}
\tikzset{block/.style={draw, thick, text width=4cm ,minimum height=1.3cm, align=center},   
line/.style={-latex}     
}  

\Crefname{algocf}{alg.}{algs.}
\Crefname{algocf}{Algorithm}{Algorithms}

\providecommand{\U}[1]{\protect\rule{.1in}{.1in}}
\newtheorem{thm}{Theorem}\Crefname{thm}{Theorem}{Theorems}
\newtheorem{lem}[thm]{Lemma}\Crefname{lem}{Lemma}{Lemmas}
\Crefname{prp}{Proposition}{Propositions}
\newtheorem{cor}[thm]{Corollary}\Crefname{cor}{Corollary}{Corollaries}
\Crefname{prb}{Problem}{Problems}
\Crefname{dfn}{Definition}{Definitions}
\Crefname{conj}{Conjecture}{Conjectures}
\Crefname{section}{Section}{Sections}
\Crefname{appendix}{Appendix}{Appendices}
\numberwithin{equation}{section}
\let\oldref\ref
\renewcommand{\ref}[1]{(\oldref{#1})}
\newcommand{\R}{\mathbb{R}}

\newcommand{\C}{\mathbb{C}}

\newcommand{\sinc}{\mathrm{sinc}}
\newcommand{\Span}{\mathrm{Span}}

\newcommand{\SQiSW}{\mathrm{SQ\lowercase{i}SW} }
\newcommand{\iSWAP}{\mathrm{\lowercase{i}SWAP} }
\newcommand{\CNOT}{\mathrm{CNOT} }
\newcommand{\CZ}{\mathrm{CZ} }

\newcommand{\SWAP}{\mathrm{SWAP} }
\newcommand{\AshN}{\mathrm{AshN} }

\begin{document}

\author[1]{Jianxin Chen}
\author[2]{Dawei Ding}
\author[3]{Weiyuan Gong}
\author[1]{Cupjin Huang}
\author[4,5]{Qi Ye}
\affil[1]{Quantum Laboratory, DAMO Academy, Bellevue, Washington 98004, USA}
\affil[2]{Quantum Laboratory, DAMO Academy, Sunnyvale, California 94085, USA}
\affil[3]{John A. Paulson School of Engineering and Applied Sciences, Harvard University, Allston, Massachusetts 02134, USA}
\affil[4]{Quantum Laboratory, DAMO Academy, Hangzhou, Zhejiang 311121, P.R.China}
\affil[5]{Institute for Interdisciplinary Information Sciences, Tsinghua University, Beijing 100084, P.R.China}
\date{}
\title{One Gate Scheme to Rule Them All: \\ Introducing a Complex Yet Reduced Instruction Set for Quantum Computing}
\maketitle

\begin{abstract}
The design and architecture of a quantum instruction set are paramount to the performance of a quantum computer. This work introduces a gate scheme for qubits with $XX+YY$ coupling that directly and efficiently realizes any two-qubit gate up to single-qubit gates. First, this scheme enables high-fidelity execution of quantum operations and achieves minimum possible gate times.  Second, since the scheme spans the entire $\textbf{SU}(4)$ group of two-qubit gates, we can use it to attain the optimal two-qubit gate count for algorithm implementation. These two advantages in synergy give rise to a quantum Complex yet Reduced Instruction Set Computer (CRISC). Though the gate scheme is compact, it supports a comprehensive array of quantum operations. This may seem paradoxical but is realizable due to the fundamental differences between quantum and classical computer architectures. 

Using our gate scheme, we observe marked improvements across various applications, including generic $n$-qubit gate synthesis, quantum volume, and qubit routing. Furthermore, the proposed scheme also realizes a gate locally equivalent to the commonly used $\CNOT$ gate with a gate time of $\frac{\pi}{2g}$, where $g$ is the two-qubit coupling. The AshN scheme is also completely impervious to $ZZ$ error, the main coherent error in transversely coupled systems, as the control parameters implementing the gates can be easily adjusted to take the $ZZ$ term into account.
\end{abstract}
\maketitle
\section{Introduction}\label{sec:intro}
With the advent of quantum computing, the development of a quantum instruction set -- an essential catalog of operations that a quantum computer can execute -- has become a critical factor in realizing a practical quantum computer. The DiVincenzo criteria~\cite{D2000}, outlining the essentials for constructing a quantum computer, encompass five computational aspects, of which three revolve around the quantum instruction set, specifically focusing on \textit{state initialization}, \textit{universal quantum operations}, and \textit{state readout}. The study and development of state initialization and readout mostly concern their physical realizations. This paper focuses on the universal quantum operations criterion, which involves not just physical realization but also the implications for algorithm implementation.

Quantum and classical instruction sets have some common qualities. For both quantum and classical computing, the instruction set is crucial as it stipulates the available operations and establishes the capabilities of a computational model. It is widely accepted that adding more instructions increases code density, that is, the inverse of the total length of instructions needed to implement algorithms, potentially improving the overall system performance. However, it is essential to recognize the fundamental differences between classical and quantum instruction sets. In a classical framework, a compact instruction set with only a few complex instructions can reduce chip size and manufacturing cost. Yet, in quantum computing, qubits, not quantum operations, are integrated within the quantum chip.\footnote{This is true at least for mainstream platforms such as superconducting circuits, trapped-ions, and neutral atoms. } Consequently, the complexity of the quantum instruction set directly impacts the complexity of qubit control, not the chip cost. From another point of view, gates sharing similar control schemes can in principle be realized with the same quantum instruction, even if they have different functionalities. This is in stark contrast to the classical setting, where different combinatorial logics need to all be hardwired onto the chip itself, necessarily increasing the chip cost.

Hence, it is possible to obtain a rich quantum instruction set but not pay the price with chip size or cost. It is therefore worthwhile to consider expanding the quantum instruction set or even including \emph{continuous} families of instructions. We will in particular focus on superconducting qubits (although our analyses would also directly apply to any other physical platform with the same Hamiltonian), for which well-known examples of realizable continuous two-qubit gate families include the fSim gate family~\cite{foxen2020demonstrating, nguyen2022programmable} and the $XY$ interaction family~\cite{abrams2020implementation}. Although these gate families cover a substantial subset of two-qubit gates, there is still a wide variety of two-qubit gates that are not included. There is thus left the tantalizing question of whether \emph{all} two-qubit gates can be realized as native instructions, and if so, whether this can be done via a universal gate scheme.

This work answers this question in the affirmative by proposing the \emph{AshN gate scheme} capable of directly realizing any two-qubit gate up to single-qubit operations. This leads to a quantum \textit{Complex yet Reduced Instruction Set Computer} (CRISC): the instruction set yields a \emph{complex} set of quantum operations, yet is \emph{reduced} in that they can all be realized with one gate scheme. This seemingly paradoxical combination is made possible because of the distinct nature of quantum computer architecture.
Specifically, the proposed quantum instruction set and the underlying gate scheme exhibit the following features:
\begin{itemize}
\item \textbf{Reduced in terms of hardware realization}: The AshN scheme can directly realize any two-qubit gate, up to single-qubit rotations, without decomposing it into multiple $\CNOT$ operations. The gate scheme to realize every gate has the same underlying physical operations, just with different control parameters. This allows for superconducting device parameter engineering \emph{that simultaneously optimizes all possible two-qubit gates modulo single-qubit gates.}

\item \textbf{Optimal gate time}: The gate time, relative to the coherence time, is a significant indicator of physical gate error, and we show that we can attain \textit{optimal} values given the two-qubit coupling strength. This enables a high-fidelity realization. 

\item \textbf{Complex in terms of expressiveness}: Our proposed gate scheme represents the first candidate to expand the quantum instruction set to fill the entire $\textbf{SU}(4)$ group, modulo single-qubit rotations. $\textbf{SU}(4)$ mathematically describes all possible noiseless evolutions of a two-qubit system.

\item \textbf{Optimal two-qubit instruction count}: By directly implementing the entire $\textbf{SU}(4)$ group modulo single qubit rotations, quantum programs can be programmed with an optimal number of two-qubit instructions. Combined with an optimal gate time for each two-qubit gate, our gate scheme improves upon previous instruction sets both in fidelity and expressiveness, significantly improving code density.


\end{itemize}

Like~\cite{foxen2020demonstrating,nguyen2022programmable}, we a priori need to calibrate each two-qubit gate in experiments.\footnote{In contrast, the scheme in~\cite{abrams2020implementation} does not need calibrating a continuous gate set, but they require multiple applications of a calibrated gate to obtain the entire continuous family. This can compromise time optimality. } As pointed out by~\cite{LMM+21}, a complex quantum instruction set may inevitably introduce significant calibration overhead, hindering its practicality for quantum computation. We argue that this may not be a fundamental barrier. We propose methods for efficiently calibrating the AshN gate set in~\Cref{sec:cal} and discuss calibrating continuous gate sets in~\Cref{sec:dis}. Note that these previous works include experimental demonstrations while AshN has yet to be implemented in hardware. However, the experiment and numerical simulation performed in~\cite{guo2018dephasing} uses the same physical setup as the AshN scheme, and their results gives good evidence that the AshN gate family can be realized with high fidelity with appropriate hardware parameter engineering. A more detailed discussion is included in~\Cref{subsec:gen_2qubit}.

\emph{Note Added}: While this paper was undergoing peer review, a related paper~\cite{wei2023native} was posted to the arXiv. 

\section{Preliminaries}
\subsection{Quantum Computing and Quantum Mechanics}
Quantum computing is a computational model motivated by physical systems governed by quantum mechanics. The different states of a quantum mechanical system is mathematically described by a complex vector space, or more specifically a \emph{Hilbert space} $\mathcal H$. For example, a \emph{qubit} state is mathematically described by a normalized vector in $\mathcal H = \mathbb{C}^2$. States of multiple qubits are described by vectors in the tensor products of their respective Hilbert spaces.

Logical gates are realized by properly engineering the dynamics of these quantum systems. The mathematical description of dynamics is given by the \emph{Hamiltonian} $H$, a Hermitian operator on $\mathcal H$. Specifically, the transformation of the state of a quantum system over a time $t$ is given by a unitary operator on $\mathcal H$, namely the matrix exponential of $H$:
\begin{align*}
    U := \exp[-i H t] : \mathcal{H} \to \mathcal H.
\end{align*}
Thus, the set of logical operators on $n$ qubits is $\textbf{SU}(2^n)$.

Hamiltonians of qubits often involve the Pauli operators:
\begin{align*}
    I:= 
    \begin{bmatrix}
    1 & 0\\
    0 & 1
    \end{bmatrix},
    X :=
    \begin{bmatrix}
    0 & 1\\
    1& 0 
    \end{bmatrix},
    Y:=
    \begin{bmatrix}
    0 & -i \\
    i& 0
    \end{bmatrix},
    Z:=
    \begin{bmatrix}
    1 & 0 \\
    0 & -1
    \end{bmatrix}.
\end{align*}
Note that $I$ is just the $2 \times 2$ identity matrix. For conciseness, we will denote tensor products of Pauli matrices as strings with alphabet $\{I,X,Y,Z\}$. For example, $XI := X \otimes I$ and $YY := Y \otimes Y$.

\subsection{The Weyl Chamber}
One useful decomposition of two-qubit unitary gates is the KAK decomposition, where an arbitrary two-qubit unitary gate $U$ is decomposed into single-qubit operations interleaved by a purely non-local interaction. 

\begin{thm}[KAK decomposition~\cite{zhang2003geometric}]
For an arbitrary $U\in \textbf{SU}(4)$, there exists a unique $\Vec{\eta}=(x,y,z) \in W\subseteq \mathbb{R}^3$, single qubit rotations $A_1,A_2,B_1,B_2\in \textbf{SU}(2)$ and a global phase $g\in\{1,i\}$ such that
\begin{align*}
U = g\cdot\left(A_1\otimes A_2\right)\exp\{i\Vec{\eta}\cdot\Vec{\Sigma}\}\left(B_1\otimes B_2\right),
\end{align*}
where $\Vec{\Sigma} \equiv [XX, YY, ZZ]$. The set
\begin{align*}
    W:=\{(x,y,z)\in\mathbb{R}^3|\pi/4\geq x\geq y\geq |z|, z\geq 0\text{ if }x=\pi/4\}
\end{align*}
is called the \emph{Weyl chamber}.  The unique vector $\Vec{\eta}\in W$ associated to $U$ is called the \emph{interaction coefficient} of $U$.
\end{thm}

\subsection{Quantum Gates in Superconducting Devices}
Superconducting qubits consist of a superconducting circuit which may involve a combination of capacitors, linear inductors, and Josephson junctions (nonlinear inductors). When superconducting, the electrons in this circuit form a superconducting condensate for which macroscopic quantities such as current and magnetic flux behave quantum mechanically according to the Hamiltonian of the circuit~\cite{yurke1984quantum,leggett1980macroscopic}. The lowest two energy levels of this quantum system (lowest eigenvalue eigenvectors of the Hamiltonian) is taken to be the ``qubit.''  In general, the circuit that implements the qubit is in composition with other circuits which is used for readout and coupling qubits together. These elements together comprise the superconducting quantum computing device.

To apply quantum gates, in general an element of the superconducting device is coupled to an external field. This external field manifests as additional terms in the Hamiltonian of the device and can be programmed to implement a desired unitary, such as $X$, on the desired qubit systems. More details can be found in~\cite{krantz2019quantum}.

\section{Physical Implementation and Hamiltonian }
\label{sec:physical_implementation}
\begin{figure}[ht]
    \centering
    \includegraphics[width=0.6\textwidth]{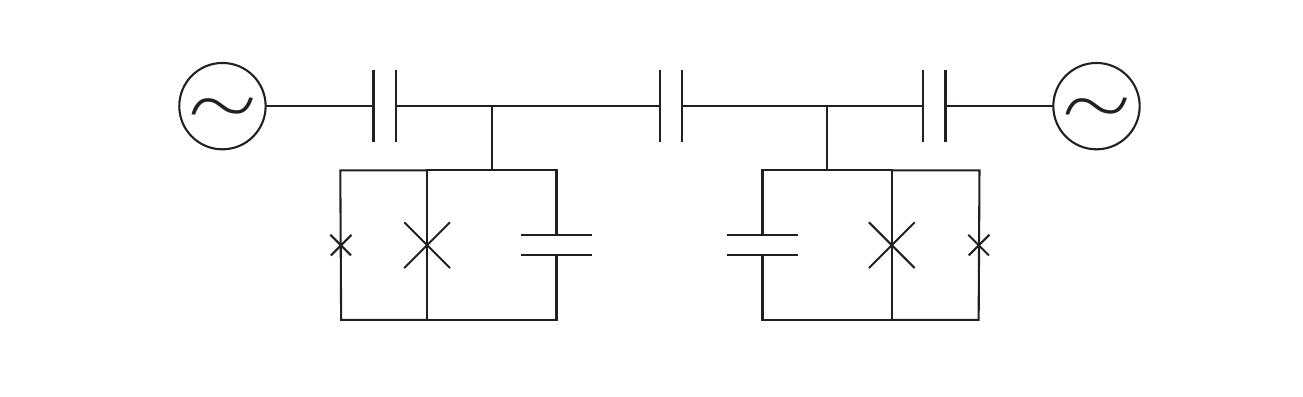}
    \caption{A pair of capacitively coupled tunable transmon qubits, each with its own microwave drive. }
    \label{fig:coupled_qubits}
\end{figure}

The $\AshN$ gate scheme is primarily motivated by considering two transversely coupled superconducting qubits in resonance while each qubit is also under the influence of a microwave drive. However, we stress that the gate scheme applies to \emph{any} physical platform that has the same Hamiltonian. This should be relatively common given that the main ingredients are just transverse coupling and single-qubit control. Going back to superconducting qubits, specific realizations could be a pair of tunable transmons or fluxonia with capacitive coupling, each with its own microwave drive. The transmon example is shown in~\Cref{fig:coupled_qubits}.

In the general setting, this setup is described by the lab frame Hamiltonian~\cite{krantz2019quantum,guo2018dephasing}
\begin{align}
\label{eq:lab_ham}
    H_L = - \frac{\omega_1}{2} ZI - \frac{\omega_2}{2} IZ + g YY + V_1(t) YI + V_2(t) IY.
\end{align}
Without loss of generality, we assume $g>0$; the case $g<0$ can be transformed to $g>0$ by changing the frame on the first qubit by a $\pi$-rotation along the $Z$ axis. To implement the $\AshN$ gate, we simply require the microwave drives to be sinusoidal with a fixed frequency and square pulse amplitude envelope $A_i(t)$:
\begin{align}
\label{eq:drive}
    V_i(t) = A_i(t) \sin(\omega_{d_i} t + \phi_i).
\end{align}
We now consider~\Cref{eq:lab_ham} in the frame rotating with each sinusoidal drive:
\begin{align*}
    H_R := i \dot U_{\text{rf}} U_{\text{rf}}^\dagger + U_{\text{rf}} H_L U_{\text{rf}}^\dagger,
\end{align*}
where
\begin{align*}
    U_\text{rf} := \exp\left[i \left(- \frac{\omega_{d_1}}{2} ZI - \frac{\omega_{d_2}}{2} IZ \right)t\right].
\end{align*}
Note that quantum operations in this frame differ from that of the frame rotating with the qubits by single-qubit phases. We compute
\begin{align*}
    H_R =& \frac{\omega_{d_1} - \omega_1}{2} ZI +\frac{\omega_{d_2} - \omega_2}{2} IZ + g (\cos(\omega_{d_1} t) Y - \sin(\omega_{d_1} t) X)\otimes (\cos(\omega_{d_2} t) Y - \sin(\omega_{d_2} t) X) \\
     +& A_1(t) \sin(\omega_{d_1}t + \phi_1) (\cos(\omega_{d_1} t) YI - \sin(\omega_{d_1} t) XI ) + A_2(t) \sin(\omega_{d_2}t + \phi_2) (\cos(\omega_{d_2} t) IY - \sin(\omega_{d_2} t) IX ).
\end{align*}
Now, we bring the qubits into resonance ($\omega_1 = \omega_2 = \omega$) and assume the drive frequencies are the same ($\omega_{d_1} = \omega_{d_2} = \omega_d $). We subsequently apply the rotating wave approximation to eliminate terms oscillating at frequencies $2 \omega_d$ to obtain\footnote{We will find that a non-negative detuning is sufficient and so $2\omega_d \geq 2\omega$. }
\begin{align}
    H_R(t) & =  \frac g 2 (XX + YY) - \frac{A_1(t)}{2} (\cos \phi_1 XI - \sin \phi_1 YI)  - \frac{A_2(t)}{2} (\cos \phi_2 IX - \sin \phi_2 IY) + \delta (ZI + IZ), \label{eq:ashn_ham}
\end{align}
where $\delta := \frac{\omega_d - \omega}{2}$ is half the drive detuning. This Hamiltonian is a more general version of the Hamiltonian in~\cite{guo2018dephasing}, where the authors also considered the same physical setup with two transversely coupled qubits. However, in their paper they considered the large amplitude limit $\vert A_1 - A_2 \vert \gg g$, while here we consider tuning the amplitudes and detunings to obtain an expressive gate set. Furthermore, they also show that their gate scheme is dephasing-insensitive, which would apply to our case when the drive amplitudes are sufficiently large compared to the detuning and qubit frequency fluctuation. Note also that unlike their scheme, we do not consider employing a phase flip in the middle of the gate to further enhance dephasing-insensitivity, although such a choice can naturally be integrated. 

\section{Spanning the Weyl Chamber}
\label{sec:ashn}
Given a system consisting of two qubits, the transverse coupling strength $g$ is usually fixed or tuned to a fixed value during an interaction, while the drive parameters $A_1,A_2,\delta$ are tunable. We now show that every two-qubit unitary can be realized via the evolution of the AshN Hamiltonian up to single-qubit gate corrections.


\subsection{The \texorpdfstring{$ZZ$}{ZZ} Coupling Problem}
\label{subsec:zz}
Before we give the explicit gate scheme, we first consider the $ZZ$ coupling problem. While decoherence is the main source of incoherent error, there can also be significant sources of coherent error when implementing the AshN scheme. In transversely coupled systems, there is often an unwanted $ZZ$ coupling between the qubits which constitutes much of the coherent gate error~\cite{foxen2020demonstrating, o2015qubit, barends2019diabatic, krinner2020benchmarking, arute2019quantum, bao2022fluxonium}. Although architectural designs such as using level engineering can mitigate the $ZZ$ coupling~\cite{sung2021realization, ding2023high}, for most superconducting devices this can still be a lingering problem. Other approaches take a ``it's not a bug, it's a feature'' stance~\cite{arute2019quantum,foxen2020demonstrating, moskalenko2021tunable, moskalenko2022high} by treating the gate implemented with $ZZ$ coupling as part of the instruction set. However, these gates can be complicated to analyze and using them for compiling other gates may require brute force search~\cite{lao2021designing}.

For the AshN gate scheme, we find that $ZZ$ coupling can be readily addressed \emph{by simply making appropriate modifications to the gate parameters}. That is, we neither treat it as a significant source of unitary error nor a necessary ad hoc change to our instruction set, but instead simply include it in the input Hamiltonian to the AshN gate scheme. We write the Hamiltonian with $ZZ$ coupling $h \in \mathbb R$ as
\begin{align}
    H_R(t) & =  \frac g 2 (XX + YY) - \frac{A_1(t)}{2} (\cos \phi_1 XI - \sin \phi_1 YI) - \frac{A_2(t)}{2} (\cos \phi_2 IX - \sin \phi_2 IY) + \delta (ZI + IZ)+ \frac{h}{2} ZZ. \label{eq:ashn_zz_ham}
\end{align}
In other words, we will prove a stronger statement than our initial claim: that $H_R$ even with nonzero $h$ spans the Weyl chamber. For technical reasons, we will require that $\vert h \vert \leq g$, which is readily satisfied in most experimental settings (see \Cref{app:ashn}).

For the AshN scheme, we assume that all microwave pulses have a square envelope.\footnote{Pulses in experiments will have nonzero rise and fall times. Like other gate schemes, this imperfection is not considered in our protocol but we believe it can be addressed with proper calibration. This is explored further in~\Cref{sec:cal}. } This makes $H_R$ time-independent. It will also be mathematically convenient to express the amplitudes in terms of symmetric and anti-symmetric combinations $\Omega_1, \Omega_2$, defined as
\begin{align}
\label{eq:square_pulse}
    A_i(t) = 
    \begin{cases}
    -2\Omega_1+(-1)^i 2\Omega_2, i\in\{1,2\},\\ 0,\text{ otherwise}.
    \end{cases}
\end{align}
Setting $\phi_1, \phi_2 =0$ and expressing $H_R$ in terms of $\Omega_1, \Omega_2$, we have 
\begin{align}
    H_R(g,h;\Omega_1,\Omega_2,\delta)
    & :=\frac g 2 (XX + YY) +\Omega_1(XI+IX)    +\Omega_2 (XI-IX) + \delta (ZI + IZ)+ \frac{h}{2} ZZ. \label{eq:ashn_zz_omega}
\end{align}
The two-qubit gate resulting from time evolving for time $\tau$ then becomes
\begin{align*}
    U(\tau;g,h;\Omega_1,\Omega_2,\delta):=\exp\{-i\cdot H_R(g,h;\Omega_1,\Omega_2,\delta)\cdot \tau\},
\end{align*}
This will be the starting point for our mathematical analyses. We emphasize again that any physical platform, not just superconducting qubits, that has the same Hamiltonian is subject to the exact same analyses.

\subsection{The Gate Scheme}
Given the Hamiltonian~\Cref{eq:ashn_zz_omega} describing our quantum computing device, our main result is the AshN gate scheme which can achieve any point in the Weyl chamber. 

\begin{algorithm}[ht!]
\caption{AshN}\label{alg:ashn}
{\small
\KwData{$(x,y,z)\in W; g\in \mathbb{R}_+;h\in[-g,g];$\\$ r\in[0,\frac{(1-|h|)\pi}{2}]$}
\KwResult{$(\tau,\Omega_1,\Omega_2,\delta)$}
$\tau_{ND}\gets 2x$\;
$\tau_{EA+}\gets 2(x+y+z)/(2-h/g)$\;
$\tau_{EA-}\gets 2(x+y-z)/(2+h/g)$\;
$\tau'_{ND}\gets \pi-2x$\;
$\tau'_{EA+}\gets 2(\pi/2-x+y-z)/(2-h/g)$\;
$\tau'_{EA-}\gets 2(\pi/2-x+y+z)/(2+h/g)$\;
$\tau_1\gets \max\{\tau_{ND}, \tau_{EA+},\tau_{EA-}\}$\;
$\tau_2\gets \max\{\tau'_{ND}, \tau'_{EA+},\tau'_{EA-}\}$\;
\eIf{$\min\{\tau_1,\tau_2\}\leq r$} {
\Return{AshN-ND-EXT($x,y,z,g,h$)};
}
{
\If{$\tau_2< \tau_1$}{
$x\gets \pi/2 - x$\;
$z\gets - z$\;
$\tau_{ND}\gets\tau'_{ND}$\;
$\tau_{EA+}\gets\tau'_{EA+}$\;
$\tau_{EA-}\gets\tau'_{EA-}$\;
}
\eIf{$\tau_{ND}\geq \max\{\tau_{EA+},\tau_{EA-}\}$}
{\Return{AshN-ND($x,y,z,g,h$)};}
{\eIf{$\tau_{EA+}\geq\tau_{EA-}$}
{\Return{AshN-EA+($x,y,z,g,h$)};}
{\Return{AshN-EA-($x,y,z,g,h$)};}
}
}
}
\end{algorithm}

\begin{algorithm}[ht!]
\caption{AshN-ND}\label{alg:ashna0}
\KwData{$(x,y,z)\in W;g\in\mathbb{R}_+;h\in[-g,g]$}
\KwResult{$(\tau,\Omega_1,\Omega_2,\delta)$}
$\tau\gets 2x$\;
$r_1\gets 2\sinc^{-1}(\frac{2\sin(y+z)}{(1-h/g)\tau})/\tau$ \Comment{$\sinc(x):=\sin(x)/x$}\;
$r_2\gets 2\sinc^{-1}(\frac{2\sin(y-z)}{(1+h/g)\tau})/\tau$
\Comment{$\sinc^{-1}:[0,1]\rightarrow [0,\pi]$}\;
$\gamma_1\gets \sqrt{r_1^2-(1-h/g)^2}/4$\;
$\gamma_2\gets \sqrt{r_2^2-(1+h/g)^2}/4$\;
\Return{$\tau/g, \gamma_1g,\gamma_2g,0$}
\end{algorithm}

\begin{algorithm}[ht!]
\caption{AshN-ND-EXT}\label{alg:ashna1}
\KwData{$(x,y,z)\in W;g\in\mathbb{R}_+;h\in[-g,g]$}
\KwResult{$(\tau,\Omega_1,\Omega_2,\delta)$}
$\tau\gets \pi-2x$\;
$r_1\gets 2\sinc^{-1}(\frac{2\sin(y-z)}{(1-h/g)\tau})/\tau$\;
$r_2\gets 2\sinc^{-1}(\frac{2\sin(y+z)}{(1+h/g)\tau})/\tau$\;
$\gamma_1\gets \sqrt{r_1^2-(1-h/g)^2}/4$\;
$\gamma_2\gets \sqrt{r_2^2-(1+h/g)^2}/4$\;
\Return{$\tau/g, \gamma_1g,\gamma_2g,0$}
\end{algorithm}

\begin{algorithm}[ht!]
\caption{AshN-EA+}\label{alg:ashnb}
\KwData{$(x,y,z)\in W;g\in\mathbb{R}_+;h\in[-g,g]$}
\KwResult{$(\tau,\Omega_1,\Omega_2,\delta)$}
$\tau\gets 2(x+y+z)/(2-h/g)$\;
$\tau'\gets (1+h/g)\tau$\;
$(x',y',z')\gets (x,y,z)+((h/g)\tau/2,(h/g)\tau/2,(h/g)\tau/2)$\;
$S\gets e^{i(y'-x'-z')}-e^{i(x'-y'-z')}-e^{i(z'-x'-y')}$\;
Find a pair $(\alpha,\beta)\in[0,1]\times [0, 2\pi/\tau]$ such that $\frac{(1-\alpha)\beta e^{i\tau'(\alpha+\beta)}}{(2\alpha+\beta)(1+\alpha+2\beta)}-\frac{(1-\alpha)(1+\alpha+\beta)e^{-i\tau'(1+\beta)}}{(1-\alpha+\beta)(1+\alpha+2\beta)}-\frac{\beta(1+\alpha+\beta) e^{-i\tau'\alpha}}{(1-\alpha+\beta)(2\alpha+\beta)}=S$\;
$\gamma\gets\sqrt{(1+\alpha+\beta)(1-\alpha)\beta}/2$\;
$d\gets\sqrt{(\alpha+\beta)\alpha(1+\beta)}/2$\;

\Return{$\tau/g, 0,(1+h/g)\gamma g,-(1+h/g)d g$};
\end{algorithm}

\begin{algorithm}[ht!]
\caption{AshN-EA-}\label{alg:ashnc}
\KwData{$(x,y,z)\in W;g\in\mathbb{R}_+;h\in[-g,g]$}
\KwResult{$(\tau,\Omega_1,\Omega_2,\delta)$}
$T',\Omega'_1,\Omega'_2,\delta'\gets \text{AshN-EA+}(x,y,-z,g,-h)$\;

\Return{$T', \Omega'_2,0,-\delta'$};
\end{algorithm}

The AshN scheme is given explicitly in~\Cref{alg:ashn} which gives the gate time $\tau$, amplitudes $\Omega_1, \Omega_2$, and detuning $\delta$ given the desired Weyl chamber coordinates $(x,y,z)$.\footnote{Note that there are conflicting definitions of the Weyl chamber coordinates in the literature and different software platforms. This may lead to a difference in sign for the coordinates computed. } Note that the single-qubit gate corrections can easily be determined using the KAK decomposition of the resulting unitary. The AshN scheme consists of four sub-schemes, corresponding to four different sectors of the Weyl chamber: No detuning (AshN-ND scheme,~\Cref{alg:ashna0}), equal amplitude (AshN-EA+,~\Cref{alg:ashnb}, and its symmetric counterpart AshN-EA-,~\Cref{alg:ashnc}), and no detuning with extended time (AshN-ND-EXT scheme,~\Cref{alg:ashna1}). 

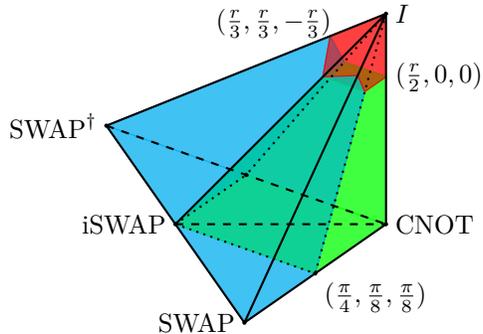
\begin{figure}
    \centering
\begin{tikzpicture}[scale=.4]
\def \tta{ 90.000000000000 } 
\def \k{    35.000000000000 } 
\def \l{     7.00000000000000 } 
\def \d{     4.00000000000000 } 
\def \h{     7.0000000000000 } 

\def \r{ 0.3}

\coordinate (I) at (0,0); 
\coordinate (Cnot) at (0,{-\h}); 
\coordinate (iSwap) at ({-\l*sin(\tta))},
                    {-\h+\l*cos(\tta)}); 
\coordinate (Swap) at ({-\l*sin(\tta)-\d*sin(\k)*sin(\tta)},{-\h+\l*cos(\tta)+\d*cos(\k)*sin(\tta)}); 
\coordinate (Swapm) at ({-\l*sin(\tta)+\d*sin(\k)*sin(\tta)},{-\h+\l*cos(\tta)-\d*cos(\k)*sin(\tta)}); 
\coordinate (SQiSW) at ({-\l*sin(\tta))/2},
                    {(-\h+\l*cos(\tta))/2});
\coordinate (mSQiSWm) at ({(-\l*sin(\tta)-\d*sin(\k)*sin(\tta))/2},{-\h+(\l*cos(\tta)+\d*cos(\k)*sin(\tta))/2});

\coordinate (m) at ({(-\l*sin(\tta)-\d*sin(\k)*sin(\tta))/2+0.4},{-\h+(\l*cos(\tta)+\d*cos(\k)*sin(\tta))/2-1.2});
\coordinate (mSQiSW) at ({(-\l*sin(\tta)+\d*sin(\k)*sin(\tta))/2},{-\h+(\l*cos(\tta)-\d*cos(\k)*sin(\tta))/2});

\coordinate (riSwap) at ({-\r*\l*sin(\tta))},
                    {-\r*\h+\r*\l*cos(\tta)}); 

\coordinate (rCnot) at (0,{-\r*\h}); 
\coordinate (rSwapm) at ({\r*2/3*(-\l*sin(\tta)-\d*sin(\k)*sin(\tta))},{\r*2/3*(-\h+\l*cos(\tta)+\d*cos(\k)*sin(\tta))}); 
\coordinate (rSwap) at ({\r*2/3*(-\l*sin(\tta)+\d*sin(\k)*sin(\tta))},{\r*2/3*(-\h+\l*cos(\tta)-\d*cos(\k)*sin(\tta))});

\coordinate (rmSQiSW) at ({\r*(-\l*sin(\tta)+\d*sin(\k)*sin(\tta))/2},{\r*(-\h+(\l*cos(\tta)-\d*cos(\k)*sin(\tta))/2)});

\coordinate (rmSQiSWm) at ({\r*(-\l*sin(\tta)-\d*sin(\k)*sin(\tta))/2},{\r*(-\h+(\l*cos(\tta)+\d*cos(\k)*sin(\tta))/2)});

\filldraw[color=red, fill opacity = 0.5]
(I)--(rSwapm) -- (rmSQiSWm) -- (rCnot);

\filldraw[color=green, fill opacity = 0.5]
(Cnot)-- (rCnot)--(rmSQiSWm)--(mSQiSWm);

\filldraw[color=cyan, fill opacity = 0.5] 
(mSQiSWm)--(rmSQiSWm)--(rSwapm)--(Swap);

\filldraw[color=green, fill opacity = 0.5]
(riSwap)--(iSwap) -- (mSQiSWm)--(rmSQiSWm);
\filldraw[color=red, fill opacity = 0.5]
(riSwap)--(rSwap) -- (rmSQiSW);
\filldraw[color=red, fill opacity = 0.5]
(riSwap)--(rSwapm) -- (rmSQiSWm);
\filldraw[color=red, fill opacity = 0.5]
(riSwap)-- (rmSQiSW) -- (rCnot) -- (rmSQiSWm);

\filldraw[color=cyan, fill opacity = 0.5]
(riSwap)--(rSwap) -- (rmSQiSW);
\filldraw[color=cyan, fill opacity = 0.5]
(riSwap)--(rSwapm) -- (rmSQiSWm);
\filldraw[color=cyan, fill opacity = 0.5]
(riSwap)--(iSwap) -- (mSQiSW)--(rmSQiSW);
\filldraw[color=cyan, fill opacity = 0.5]
(riSwap)--(iSwap) -- (mSQiSWm)--(rmSQiSWm);
\filldraw[color=cyan, fill opacity = 0.5]
(iSwap)--(Swap) -- (mSQiSWm);
\filldraw[color=cyan, fill opacity = 0.5]
(iSwap)--(Swapm) -- (mSQiSW);

\filldraw[color=green, fill opacity = 0.5]
(riSwap)-- (rmSQiSW) -- (rCnot) -- (rmSQiSWm);

\filldraw[color=green, fill opacity = 0.5]
(riSwap)--(iSwap) -- (mSQiSW)--(rmSQiSW);
\filldraw[color=green, fill opacity = 0.5]
(iSwap)-- (mSQiSWm)--(Cnot)--(mSQiSW);
\filldraw[color=green, fill opacity = 0.5]
(Cnot)-- (rCnot)--(rmSQiSW)--(mSQiSW);

\filldraw[color=cyan, fill opacity = 0.5] 
(mSQiSW)--(rmSQiSW)--(rSwap)--(Swapm);
\filldraw[color=cyan, fill opacity = 0.5] 
(Swap) -- (rSwapm)--(riSwap)--(rSwap)--(Swapm);
\filldraw[color=red, fill opacity = 0.5] 
(I)-- (rSwapm) -- (riSwap) -- (rSwap);
\filldraw[color=red, fill opacity = 0.5]
(I)--(rSwap) -- (rmSQiSW) -- (rCnot);

                    
\draw[-,thick] (Cnot) --  (Swapm)
                        (I) --  (Swap)
                        (I) -- (Swapm)
                        (I) -- (Cnot)
                        (Swap) --  (Swapm)
                        (I) --  (iSwap);

\draw[dashed,thick] (iSwap) --  (Cnot)
                        (Swap)  -- (Cnot);

\draw[dotted, thick] (iSwap) --  (mSQiSW)
                        (iSwap)  -- (mSQiSWm)
                        (I) -- (mSQiSW)
                        (I) -- (mSQiSWm);

\fill[black]  (I) circle [radius=2pt]; 
\fill[black]    (iSwap) circle [radius=2pt]; 
\fill[black]  (Cnot) circle [radius=2pt]; 
\fill[black] (Swap) circle [radius=2pt];
\fill[black] (Swapm) circle [radius=2pt];
\fill[black] (mSQiSW) circle [radius=2pt];
\fill[black] (mSQiSWm) circle [radius=2pt];

\draw (I) node [right]           {$I$}
      (Cnot) node [right]     {$\CNOT$}
      (Swap) node [left]           {$\SWAP^\dag$}
      (Swapm)  node [left]       {$\SWAP$}
      (iSwap)  node [left]            {$\iSWAP$}
      (rCnot) node [right] {$(\frac r 2,0,0)$}
      (mSQiSW) node [below right] {$(\frac\pi4,\frac\pi8,\frac\pi8)$}
      (rSwapm) node [above left=-0.2cm] {$(\frac{r}{3}, \frac{r}3, -\frac{r}3)$};
\end{tikzpicture}
\caption{Illustration of the AshN gate scheme with no $ZZ$ coupling. Depending on the Weyl chamber coordinate and the cutoff value $r$, a gate can either be implemented in optimal time using the AshN-ND scheme (green) or the AshN-EA+/- scheme (blue), or with slightly longer times with assurance of finite drive amplitudes and detunings using the AshN-ND-EXT scheme (red).}
\label{fig:ashn}
\end{figure}

\begin{figure}
\centering
\subfloat[$h=0.2g$.\label{fig:0.2g}]{
\begin{tikzpicture}[scale=.3]
\def \tta{ 90.000000000000 } 
\def \k{    35.000000000000 } 
\def \l{     7.00000000000000 } 
\def \d{     4.00000000000000 } 
\def \h{     7.0000000000000 } 

\def \r{ 0.3}
\def \g{ 0.2}

\coordinate (I) at (0,0); 
\coordinate (Cnot) at (0,{-\h}); 
\coordinate (iSwap) at ({-\l*sin(\tta)+\d*\g*sin(\k)*sin(\tta))},
                    {-\h+\l*cos(\tta)-\d*\g*cos(\k)*sin(\tta)}); 
\coordinate (mSQiSWm) at ({(-\l*(1-\g)*sin(\tta)-\d*(1-\g)*sin(\k)*sin(\tta))/2},{-\h+(\l*(1-\g)*cos(\tta)+\d*(1-\g)*cos(\k)*sin(\tta))/2});

\coordinate (mSQiSW) at ({(-\l*(1+\g)*sin(\tta)+\d*(1+\g)*sin(\k)*sin(\tta))/2},{-\h+(\l*(1+\g)*cos(\tta)-\d*(1+\g)*cos(\k)*sin(\tta))/2});

\coordinate (Swap) at ({-\l*sin(\tta)-\d*sin(\k)*sin(\tta)},{-\h+\l*cos(\tta)+\d*cos(\k)*sin(\tta)}); 
\coordinate (Swapm) at ({-\l*sin(\tta)+\d*sin(\k)*sin(\tta)},{-\h+\l*cos(\tta)-\d*cos(\k)*sin(\tta)}); 

\coordinate (riSwap) at ({\r*(-\l*sin(\tta)+\d*\g*sin(\k)*sin(\tta)))},
                    {\r*(-\h+\l*cos(\tta)-\d*\g*cos(\k)*sin(\tta))}); 

\coordinate (rCnot) at (0,{-\r*\h}); 
\coordinate (rSwapm) at ({\r*2/3*(1-\g/2)*(-\l*sin(\tta)-\d*sin(\k)*sin(\tta))},{\r*2/3*(1-\g/2)*(-\h+\l*cos(\tta)+\d*cos(\k)*sin(\tta))}); 
\coordinate (rSwap) at ({\r*2/3*(1+\g/2)*(-\l*sin(\tta)+\d*sin(\k)*sin(\tta))},{\r*2/3*(1+\g/2)*(-\h+\l*cos(\tta)-\d*cos(\k)*sin(\tta))});

\coordinate (rmSQiSW) at ({\r*((-\l*(1+\g)*sin(\tta)+\d*(1+\g)*sin(\k)*sin(\tta))/2)},{\r*(-\h+(\l*(1+\g)*cos(\tta)-\d*(1+\g)*cos(\k)*sin(\tta))/2)});

\coordinate (rmSQiSWm) at ({\r*((-\l*(1-\g)*sin(\tta)-\d*(1-\g)*sin(\k)*sin(\tta))/2)},{\r*(-\h+(\l*(1-\g)*cos(\tta)+\d*(1-\g)*cos(\k)*sin(\tta))/2)});

\coordinate (niSwap) at ({-\l*sin(\tta)-\d*\g*sin(\k)*sin(\tta))},
                    {-\h+\l*cos(\tta)+\d*\g*cos(\k)*sin(\tta)}); 
\coordinate (nmSQiSWm) at ({(-\l*(1+\g)*sin(\tta)-\d*(1+\g)*sin(\k)*sin(\tta))/2},{-\h+(\l*(1+\g)*cos(\tta)+\d*(1+\g)*cos(\k)*sin(\tta))/2});

\coordinate (nSwapm) at ({(-\l*sin(\tta)-\d*sin(\k)*sin(\tta)) *(2*(1+\g)/(3+\g))},{(-\h+\l*cos(\tta)+\d*cos(\k)*sin(\tta)) *(2*(1+\g)/(3+\g))}); 
\coordinate (tiSwap) at ({-\l*sin(\tta)},
                    {-\h+\l*cos(\tta)}); 
\coordinate (tiSwapr) at ({-\l*sin(\tta) * (1-\g)},
                    {-\h+\l*cos(\tta) * (1-\g)}); 
\coordinate (nnSwapm) at ({(-\l*sin(\tta)-\d*sin(\k)*sin(\tta)) *(2*(2-\g)/(5-\g))},{(-\h+\l*cos(\tta)+\d*cos(\k)*sin(\tta)) *(2*(2-\g)/(5-\g))}); 
\coordinate(nxy) at ({-\l*sin(\tta) * (3-2*\g)/(3-\g)+\d*(-\g/(3-\g))*sin(\k)*sin(\tta))},
                    {(-\h+\l*cos(\tta)) * (3-2*\g)/(3-\g) -\d*(-\g/(3-\g))*cos(\k)*sin(\tta)});

\filldraw[color=green, fill opacity = 0.5]
(nxy)-- (nnSwapm) -- (mSQiSWm) -- (tiSwapr);
\filldraw[color=green, fill opacity = 0.5]
(nxy)-- (nnSwapm) -- (nSwapm) -- (niSwap);
\filldraw[color=green, fill opacity = 0.5]
(nxy)-- (tiSwapr) -- (niSwap);

\filldraw[color=red, fill opacity = 0.5]
(I)--(rSwapm) -- (rmSQiSWm) -- (rCnot);

\filldraw[color=green, fill opacity = 0.5]
(Cnot)-- (rCnot)--(rmSQiSWm)--(mSQiSWm);

\filldraw[color=cyan, fill opacity = 0.5] 
(mSQiSWm)--(rmSQiSWm)--(rSwapm)--(Swap);

\filldraw[color=green, fill opacity = 0.5]
(riSwap)--(iSwap) -- (mSQiSWm)--(rmSQiSWm);
\filldraw[color=red, fill opacity = 0.5]
(riSwap)--(rSwap) -- (rmSQiSW);
\filldraw[color=red, fill opacity = 0.5]
(riSwap)--(rSwapm) -- (rmSQiSWm);
\filldraw[color=red, fill opacity = 0.5]
(riSwap)-- (rmSQiSW) -- (rCnot) -- (rmSQiSWm);

\filldraw[color=cyan, fill opacity = 0.5]
(riSwap)--(rSwap) -- (rmSQiSW);
\filldraw[color=cyan, fill opacity = 0.5]
(riSwap)--(rSwapm) -- (rmSQiSWm);
\filldraw[color=cyan, fill opacity = 0.5]
(riSwap)--(iSwap) -- (mSQiSW)--(rmSQiSW);
\filldraw[color=cyan, fill opacity = 0.5]
(riSwap)--(iSwap) -- (mSQiSWm)--(rmSQiSWm);
\filldraw[color=cyan, fill opacity = 0.5]
(iSwap)--(Swap) -- (mSQiSWm);
\filldraw[color=cyan, fill opacity = 0.5]
(iSwap)--(Swapm) -- (mSQiSW);

\filldraw[color=green, fill opacity = 0.5]
(riSwap)-- (rmSQiSW) -- (rCnot) -- (rmSQiSWm);

\filldraw[color=green, fill opacity = 0.5]
(riSwap)--(iSwap) -- (mSQiSW)--(rmSQiSW);
\filldraw[color=green, fill opacity = 0.5]
(iSwap)-- (mSQiSWm)--(Cnot)--(mSQiSW);
\filldraw[color=green, fill opacity = 0.5]
(Cnot)-- (rCnot)--(rmSQiSW)--(mSQiSW);

\filldraw[color=cyan, fill opacity = 0.5] 
(mSQiSW)--(rmSQiSW)--(rSwap)--(Swapm);
\filldraw[color=cyan, fill opacity = 0.5] 
(Swap) -- (rSwapm)--(riSwap)--(rSwap)--(Swapm);
\filldraw[color=red, fill opacity = 0.5] 
(I)-- (rSwapm) -- (riSwap) -- (rSwap);
\filldraw[color=red, fill opacity = 0.5]
(I)--(rSwap) -- (rmSQiSW) -- (rCnot);

                    
\draw[-,thick] (Cnot) --  (Swapm)
                        (I) --  (Swap)
                        (I) -- (Swapm)
                        (I) -- (Cnot)
                        (Swap) --  (Swapm)
                        (I) --  (iSwap);

\draw[dashed,thick] (iSwap) --  (Cnot)
                        (Swap)  -- (Cnot);

\draw[dotted, thick] (iSwap) --  (mSQiSW)
                        (iSwap)  -- (mSQiSWm)
                        (I) -- (mSQiSW)
                        (I) -- (mSQiSWm)
                        (niSwap) -- (nmSQiSWm)
                        (niSwap) -- (nSwapm)
                        (nSwapm) -- (nmSQiSWm)
                        (tiSwap) -- (niSwap)
                        (niSwap) -- (tiSwapr)
                        (tiSwapr) -- (tiSwap)
                        (nxy) -- (niSwap)
                        (nxy) -- (tiSwap)
                        (nxy) -- (tiSwapr)
                        (nxy) -- (nnSwapm)
                        (nnSwapm) -- (mSQiSWm);

\fill[black]  (I) circle [radius=2pt]; 
\fill[black]    (iSwap) circle [radius=2pt]; 
\fill[black]  (Cnot) circle [radius=2pt]; 
\fill[black] (Swap) circle [radius=2pt];
\fill[black] (Swapm) circle [radius=2pt];
\fill[black] (mSQiSW) circle [radius=2pt];
\fill[black] (mSQiSWm) circle [radius=2pt];

\end{tikzpicture}
}~
\subfloat[$h=0.4g$.\label{fig:0.4g}]{
\begin{tikzpicture}[scale=.3]
\def \tta{ 90.000000000000 } 
\def \k{    35.000000000000 } 
\def \l{     7.00000000000000 } 
\def \d{     4.00000000000000 } 
\def \h{     7.0000000000000 } 

\def \r{ 0.3}
\def \g{ 0.4}

\coordinate (I) at (0,0); 
\coordinate (Cnot) at (0,{-\h}); 
\coordinate (iSwap) at ({-\l*sin(\tta)+\d*\g*sin(\k)*sin(\tta))},
                    {-\h+\l*cos(\tta)-\d*\g*cos(\k)*sin(\tta)}); 
\coordinate (Swap) at ({-\l*sin(\tta)-\d*sin(\k)*sin(\tta)},{-\h+\l*cos(\tta)+\d*cos(\k)*sin(\tta)}); 
\coordinate (Swapm) at ({-\l*sin(\tta)+\d*sin(\k)*sin(\tta)},{-\h+\l*cos(\tta)-\d*cos(\k)*sin(\tta)}); 
\coordinate (mSQiSWm) at ({(-\l*(1-\g)*sin(\tta)-\d*(1-\g)*sin(\k)*sin(\tta))/2},{-\h+(\l*(1-\g)*cos(\tta)+\d*(1-\g)*cos(\k)*sin(\tta))/2});

\coordinate (mSQiSW) at ({(-\l*(1+\g)*sin(\tta)+\d*(1+\g)*sin(\k)*sin(\tta))/2},{-\h+(\l*(1+\g)*cos(\tta)-\d*(1+\g)*cos(\k)*sin(\tta))/2});

\coordinate (riSwap) at ({\r*(-\l*sin(\tta)+\d*\g*sin(\k)*sin(\tta)))},
                    {\r*(-\h+\l*cos(\tta)-\d*\g*cos(\k)*sin(\tta))}); 

\coordinate (rCnot) at (0,{-\r*\h}); 
\coordinate (rSwapm) at ({\r*2/3*(1-\g/2)*(-\l*sin(\tta)-\d*sin(\k)*sin(\tta))},{\r*2/3*(1-\g/2)*(-\h+\l*cos(\tta)+\d*cos(\k)*sin(\tta))}); 
\coordinate (rSwap) at ({\r*2/3*(1+\g/2)*(-\l*sin(\tta)+\d*sin(\k)*sin(\tta))},{\r*2/3*(1+\g/2)*(-\h+\l*cos(\tta)-\d*cos(\k)*sin(\tta))});

\coordinate (rmSQiSW) at ({\r*((-\l*(1+\g)*sin(\tta)+\d*(1+\g)*sin(\k)*sin(\tta))/2)},{\r*(-\h+(\l*(1+\g)*cos(\tta)-\d*(1+\g)*cos(\k)*sin(\tta))/2)});

\coordinate (rmSQiSWm) at ({\r*((-\l*(1-\g)*sin(\tta)-\d*(1-\g)*sin(\k)*sin(\tta))/2)},{\r*(-\h+(\l*(1-\g)*cos(\tta)+\d*(1-\g)*cos(\k)*sin(\tta))/2)});

\coordinate (niSwap) at ({-\l*sin(\tta)-\d*\g*sin(\k)*sin(\tta))},
                    {-\h+\l*cos(\tta)+\d*\g*cos(\k)*sin(\tta)}); 
\coordinate (nmSQiSWm) at ({(-\l*(1+\g)*sin(\tta)-\d*(1+\g)*sin(\k)*sin(\tta))/2},{-\h+(\l*(1+\g)*cos(\tta)+\d*(1+\g)*cos(\k)*sin(\tta))/2});

\coordinate (nSwapm) at ({(-\l*sin(\tta)-\d*sin(\k)*sin(\tta)) *(2*(1+\g)/(3+\g))},{(-\h+\l*cos(\tta)+\d*cos(\k)*sin(\tta)) *(2*(1+\g)/(3+\g))}); 
\coordinate (tiSwap) at ({-\l*sin(\tta)},
                    {-\h+\l*cos(\tta)}); 
\coordinate (tiSwapr) at ({-\l*sin(\tta) * (1-\g)},
                    {-\h+\l*cos(\tta) * (1-\g)}); 
\coordinate (nnSwapm) at ({(-\l*sin(\tta)-\d*sin(\k)*sin(\tta)) *(2*(2-\g)/(5-\g))},{(-\h+\l*cos(\tta)+\d*cos(\k)*sin(\tta)) *(2*(2-\g)/(5-\g))}); 
\coordinate(nxy) at ({-\l*sin(\tta) * (3-2*\g)/(3-\g)+\d*(-\g/(3-\g))*sin(\k)*sin(\tta))},
                    {(-\h+\l*cos(\tta)) * (3-2*\g)/(3-\g) -\d*(-\g/(3-\g))*cos(\k)*sin(\tta)});

\filldraw[color=green, fill opacity = 0.5]
(nxy)-- (nnSwapm) -- (mSQiSWm) -- (tiSwapr);
\filldraw[color=green, fill opacity = 0.5]
(nxy)-- (nnSwapm) -- (nSwapm) -- (niSwap);
\filldraw[color=green, fill opacity = 0.5]
(nxy)-- (tiSwapr) -- (niSwap);

\filldraw[color=red, fill opacity = 0.5]
(I)--(rSwapm) -- (rmSQiSWm) -- (rCnot);

\filldraw[color=green, fill opacity = 0.5]
(Cnot)-- (rCnot)--(rmSQiSWm)--(mSQiSWm);

\filldraw[color=cyan, fill opacity = 0.5] 
(mSQiSWm)--(rmSQiSWm)--(rSwapm)--(Swap);

\filldraw[color=green, fill opacity = 0.5]
(riSwap)--(iSwap) -- (mSQiSWm)--(rmSQiSWm);
\filldraw[color=red, fill opacity = 0.5]
(riSwap)--(rSwap) -- (rmSQiSW);
\filldraw[color=red, fill opacity = 0.5]
(riSwap)--(rSwapm) -- (rmSQiSWm);
\filldraw[color=red, fill opacity = 0.5]
(riSwap)-- (rmSQiSW) -- (rCnot) -- (rmSQiSWm);

\filldraw[color=cyan, fill opacity = 0.5]
(riSwap)--(rSwap) -- (rmSQiSW);
\filldraw[color=cyan, fill opacity = 0.5]
(riSwap)--(rSwapm) -- (rmSQiSWm);
\filldraw[color=cyan, fill opacity = 0.5]
(riSwap)--(iSwap) -- (mSQiSW)--(rmSQiSW);
\filldraw[color=cyan, fill opacity = 0.5]
(riSwap)--(iSwap) -- (mSQiSWm)--(rmSQiSWm);
\filldraw[color=cyan, fill opacity = 0.5]
(iSwap)--(Swap) -- (mSQiSWm);
\filldraw[color=cyan, fill opacity = 0.5]
(iSwap)--(Swapm) -- (mSQiSW);

\filldraw[color=green, fill opacity = 0.5]
(riSwap)-- (rmSQiSW) -- (rCnot) -- (rmSQiSWm);

\filldraw[color=green, fill opacity = 0.5]
(riSwap)--(iSwap) -- (mSQiSW)--(rmSQiSW);
\filldraw[color=green, fill opacity = 0.5]
(iSwap)-- (mSQiSWm)--(Cnot)--(mSQiSW);
\filldraw[color=green, fill opacity = 0.5]
(Cnot)-- (rCnot)--(rmSQiSW)--(mSQiSW);

\filldraw[color=cyan, fill opacity = 0.5] 
(mSQiSW)--(rmSQiSW)--(rSwap)--(Swapm);
\filldraw[color=cyan, fill opacity = 0.5] 
(Swap) -- (rSwapm)--(riSwap)--(rSwap)--(Swapm);
\filldraw[color=red, fill opacity = 0.5] 
(I)-- (rSwapm) -- (riSwap) -- (rSwap);
\filldraw[color=red, fill opacity = 0.5]
(I)--(rSwap) -- (rmSQiSW) -- (rCnot);

                    
\draw[-,thick] (Cnot) --  (Swapm)
                        (I) --  (Swap)
                        (I) -- (Swapm)
                        (I) -- (Cnot)
                        (Swap) --  (Swapm)
                        (I) --  (iSwap);

\draw[dashed,thick] (iSwap) --  (Cnot)
                        (Swap)  -- (Cnot);

\draw[dotted, thick] (iSwap) --  (mSQiSW)
                        (iSwap)  -- (mSQiSWm)
                        (I) -- (mSQiSW)
                        (I) -- (mSQiSWm)
                        (niSwap) -- (nmSQiSWm)
                        (niSwap) -- (nSwapm)
                        (nSwapm) -- (nmSQiSWm)
                        (tiSwap) -- (niSwap)
                        (niSwap) -- (tiSwapr)
                        (tiSwapr) -- (tiSwap)
                        (nxy) -- (niSwap)
                        (nxy) -- (tiSwap)
                        (nxy) -- (tiSwapr)
                        (nxy) -- (nnSwapm)
                        (nnSwapm) -- (mSQiSWm);

\fill[black]  (I) circle [radius=2pt]; 
\fill[black]    (iSwap) circle [radius=2pt]; 
\fill[black]  (Cnot) circle [radius=2pt]; 
\fill[black] (Swap) circle [radius=2pt];
\fill[black] (Swapm) circle [radius=2pt];
\fill[black] (mSQiSW) circle [radius=2pt];
\fill[black] (mSQiSWm) circle [radius=2pt];

\end{tikzpicture}
}~
\subfloat[$h=0.8g$.\label{fig:0.8g}]{
\begin{tikzpicture}[scale=.3]
\def \tta{ 90.000000000000 } 
\def \k{    35.000000000000 } 
\def \l{     7.00000000000000 } 
\def \d{     4.00000000000000 } 
\def \h{     7.0000000000000 } 

\def \r{ 0.3}
\def \g{ 0.8}

\coordinate (I) at (0,0); 
\coordinate (Cnot) at (0,{-\h}); 
\coordinate (iSwap) at ({-\l*sin(\tta)+\d*\g*sin(\k)*sin(\tta))},
                    {-\h+\l*cos(\tta)-\d*\g*cos(\k)*sin(\tta)}); 
\coordinate (Swap) at ({-\l*sin(\tta)-\d*sin(\k)*sin(\tta)},{-\h+\l*cos(\tta)+\d*cos(\k)*sin(\tta)}); 
\coordinate (Swapm) at ({-\l*sin(\tta)+\d*sin(\k)*sin(\tta)},{-\h+\l*cos(\tta)-\d*cos(\k)*sin(\tta)}); 
\coordinate (mSQiSWm) at ({(-\l*(1-\g)*sin(\tta)-\d*(1-\g)*sin(\k)*sin(\tta))/2},{-\h+(\l*(1-\g)*cos(\tta)+\d*(1-\g)*cos(\k)*sin(\tta))/2});

\coordinate (mSQiSW) at ({(-\l*(1+\g)*sin(\tta)+\d*(1+\g)*sin(\k)*sin(\tta))/2},{-\h+(\l*(1+\g)*cos(\tta)-\d*(1+\g)*cos(\k)*sin(\tta))/2});

\coordinate (riSwap) at ({\r*(-\l*sin(\tta)+\d*\g*sin(\k)*sin(\tta)))},
                    {\r*(-\h+\l*cos(\tta)-\d*\g*cos(\k)*sin(\tta))}); 

\coordinate (rCnot) at (0,{-\r*\h}); 
\coordinate (rSwapm) at ({\r*2/3*(1-\g/2)*(-\l*sin(\tta)-\d*sin(\k)*sin(\tta))},{\r*2/3*(1-\g/2)*(-\h+\l*cos(\tta)+\d*cos(\k)*sin(\tta))}); 
\coordinate (rSwap) at ({\r*2/3*(1+\g/2)*(-\l*sin(\tta)+\d*sin(\k)*sin(\tta))},{\r*2/3*(1+\g/2)*(-\h+\l*cos(\tta)-\d*cos(\k)*sin(\tta))});

\coordinate (rmSQiSW) at ({\r*((-\l*(1+\g)*sin(\tta)+\d*(1+\g)*sin(\k)*sin(\tta))/2)},{\r*(-\h+(\l*(1+\g)*cos(\tta)-\d*(1+\g)*cos(\k)*sin(\tta))/2)});

\coordinate (rmSQiSWm) at ({\r*((-\l*(1-\g)*sin(\tta)-\d*(1-\g)*sin(\k)*sin(\tta))/2)},{\r*(-\h+(\l*(1-\g)*cos(\tta)+\d*(1-\g)*cos(\k)*sin(\tta))/2)});

\coordinate (niSwap) at ({-\l*sin(\tta)-\d*\g*sin(\k)*sin(\tta))},
                    {-\h+\l*cos(\tta)+\d*\g*cos(\k)*sin(\tta)}); 
\coordinate (nmSQiSWm) at ({(-\l*(1+\g)*sin(\tta)-\d*(1+\g)*sin(\k)*sin(\tta))/2},{-\h+(\l*(1+\g)*cos(\tta)+\d*(1+\g)*cos(\k)*sin(\tta))/2});

\coordinate (nSwapm) at ({(-\l*sin(\tta)-\d*sin(\k)*sin(\tta)) *(2*(1+\g)/(3+\g))},{(-\h+\l*cos(\tta)+\d*cos(\k)*sin(\tta)) *(2*(1+\g)/(3+\g))}); 
\coordinate (tiSwap) at ({-\l*sin(\tta)},
                    {-\h+\l*cos(\tta)}); 
\coordinate (tiSwapr) at ({-\l*sin(\tta) * (1-\g)},
                    {-\h+\l*cos(\tta) * (1-\g)}); 
\coordinate (nnSwapm) at ({(-\l*sin(\tta)-\d*sin(\k)*sin(\tta)) *(2*(2-\g)/(5-\g))},{(-\h+\l*cos(\tta)+\d*cos(\k)*sin(\tta)) *(2*(2-\g)/(5-\g))}); 
\coordinate(nxy) at ({-\l*sin(\tta) * (3-2*\g)/(3-\g)+\d*(-\g/(3-\g))*sin(\k)*sin(\tta))},
                    {(-\h+\l*cos(\tta)) * (3-2*\g)/(3-\g) -\d*(-\g/(3-\g))*cos(\k)*sin(\tta)});

\filldraw[color=green, fill opacity = 0.5]
(nxy)-- (nnSwapm) -- (mSQiSWm) -- (tiSwapr);
\filldraw[color=green, fill opacity = 0.5]
(nxy)-- (nnSwapm) -- (nSwapm) -- (niSwap);
\filldraw[color=green, fill opacity = 0.5]
(nxy)-- (tiSwapr) -- (niSwap);

\filldraw[color=red, fill opacity = 0.5]
(I)--(rSwapm) -- (rmSQiSWm) -- (rCnot);

\filldraw[color=green, fill opacity = 0.5]
(Cnot)-- (rCnot)--(rmSQiSWm)--(mSQiSWm);

\filldraw[color=cyan, fill opacity = 0.5] 
(mSQiSWm)--(rmSQiSWm)--(rSwapm)--(Swap);

\filldraw[color=green, fill opacity = 0.5]
(riSwap)--(iSwap) -- (mSQiSWm)--(rmSQiSWm);
\filldraw[color=red, fill opacity = 0.5]
(riSwap)--(rSwap) -- (rmSQiSW);
\filldraw[color=red, fill opacity = 0.5]
(riSwap)--(rSwapm) -- (rmSQiSWm);
\filldraw[color=red, fill opacity = 0.5]
(riSwap)-- (rmSQiSW) -- (rCnot) -- (rmSQiSWm);

\filldraw[color=cyan, fill opacity = 0.5]
(riSwap)--(rSwap) -- (rmSQiSW);
\filldraw[color=cyan, fill opacity = 0.5]
(riSwap)--(rSwapm) -- (rmSQiSWm);
\filldraw[color=cyan, fill opacity = 0.5]
(riSwap)--(iSwap) -- (mSQiSW)--(rmSQiSW);
\filldraw[color=cyan, fill opacity = 0.5]
(riSwap)--(iSwap) -- (mSQiSWm)--(rmSQiSWm);
\filldraw[color=cyan, fill opacity = 0.5]
(iSwap)--(Swap) -- (mSQiSWm);
\filldraw[color=cyan, fill opacity = 0.5]
(iSwap)--(Swapm) -- (mSQiSW);

\filldraw[color=green, fill opacity = 0.5]
(riSwap)-- (rmSQiSW) -- (rCnot) -- (rmSQiSWm);

\filldraw[color=green, fill opacity = 0.5]
(riSwap)--(iSwap) -- (mSQiSW)--(rmSQiSW);
\filldraw[color=green, fill opacity = 0.5]
(iSwap)-- (mSQiSWm)--(Cnot)--(mSQiSW);
\filldraw[color=green, fill opacity = 0.5]
(Cnot)-- (rCnot)--(rmSQiSW)--(mSQiSW);

\filldraw[color=cyan, fill opacity = 0.5] 
(mSQiSW)--(rmSQiSW)--(rSwap)--(Swapm);
\filldraw[color=cyan, fill opacity = 0.5] 
(Swap) -- (rSwapm)--(riSwap)--(rSwap)--(Swapm);
\filldraw[color=red, fill opacity = 0.5] 
(I)-- (rSwapm) -- (riSwap) -- (rSwap);
\filldraw[color=red, fill opacity = 0.5]
(I)--(rSwap) -- (rmSQiSW) -- (rCnot);

                    
\draw[-,thick] (Cnot) --  (Swapm)
                        (I) --  (Swap)
                        (I) -- (Swapm)
                        (I) -- (Cnot)
                        (Swap) --  (Swapm)
                        (I) --  (iSwap);

\draw[dashed,thick] (iSwap) --  (Cnot)
                        (Swap)  -- (Cnot);

\draw[dotted, thick] (iSwap) --  (mSQiSW)
                        (iSwap)  -- (mSQiSWm)
                        (I) -- (mSQiSW)
                        (I) -- (mSQiSWm)
                        (niSwap) -- (nmSQiSWm)
                        (niSwap) -- (nSwapm)
                        (nSwapm) -- (nmSQiSWm)
                        (tiSwap) -- (niSwap)
                        (niSwap) -- (tiSwapr)
                        (tiSwapr) -- (tiSwap)
                        (nxy) -- (niSwap)
                        (nxy) -- (tiSwap)
                        (nxy) -- (tiSwapr)
                        (nxy) -- (nnSwapm)
                        (nnSwapm) -- (mSQiSWm);

\fill[black]  (I) circle [radius=2pt]; 
\fill[black]    (iSwap) circle [radius=2pt]; 
\fill[black]  (Cnot) circle [radius=2pt]; 
\fill[black] (Swap) circle [radius=2pt];
\fill[black] (Swapm) circle [radius=2pt];
\fill[black] (mSQiSW) circle [radius=2pt];
\fill[black] (mSQiSWm) circle [radius=2pt];

\end{tikzpicture}
}
\caption{Illustration of the AshN gate scheme in the presence of $ZZ$ coupling. The AshN-ND (green), AshN-EA+/- (blue) and AshN-ND-EXT (red) schemes can all be appropriately modified for $\vert h\vert \leq g$. The sectors covered by each modified scheme is also different depending on the ratio $h/g$. Note that when $h\neq 0$ the Weyl chamber is divided into seven regions instead of four, corresponding to different ways the optimal time is achieved.}
\label{fig:ashn_zz}
\end{figure}
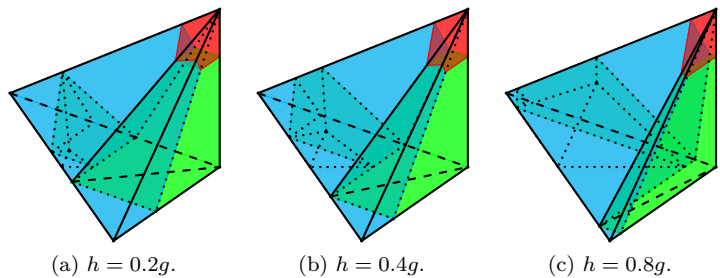

The AshN-ND and the AshN-EA+/- schemes can be used to generate an arbitrary Weyl chamber coordinate in optimal time (this will be defined in~\Cref{subsec:time_op}), but the schemes may require unbounded drive amplitudes. To address this problem, we introduce the AshN-ND-EXT scheme that requires slightly more time but with bounded detuning and amplitude. We can introduce a configurable cutoff hyper-parameter $r\in[0,(1-|h|)\pi/2]$ to switch between AshN-ND-EXT or AshN-ND / AshN-EA, achieving a trade-off between gate time and detuning or drive amplitude. The parameter $r$ has an intuitive geometric interpretation as a bounding the Weyl chamber coordinate, as shown in~\Cref{fig:ashn} for the $h=0$ case and~\Cref{fig:ashn_zz} for the $h\neq 0$ case. In~\Cref{app:ashn} we prove that by introducing any $r \in (0,(1-\vert h \vert)\pi/2]$, we can get a uniform upper bound on the amplitudes and detunings. For $h=0$ we can prove an explicit bound on the amplitudes and detunings in terms of $r$:
\begin{align}
\label{eq:r_cutoff}
  \vert A_1\vert/2, \vert A_2 \vert/2, \vert \delta \vert \leq g\left(\frac{\pi}r+\frac12\right).
\end{align}


\subsection{Time Optimality}
\label{subsec:time_op}
Decoherence is often one of the main sources of gate error, and so an important property to establish for a gate scheme is the required gate time. 
In~\cite{hammerer2002characterization}, the authors define the \emph{interaction cost} of interaction coefficients $\vec \eta =(x,y,z) \in W$ with respect to a two-qubit Hamiltonian $H$ as the minimal total Hamiltonian evolution time needed for a sequence of time evolutions via $H$ interspersed with arbitrary single-qubit gates to achieve a gate with interaction coefficients $\vec \eta$. According to their Result 2,  the interaction cost is given by the minimal time $\tau \geq 0$ such that either
\begin{align}
\label{eq:major1}
    \vec \eta \prec_s \vec\alpha\tau 
\end{align}
or 
\begin{align}
\label{eq:major2}
    \vec \eta - (\pi/2,0,0) \prec_s \vec \alpha \tau,
\end{align}
where $\prec_s$ is a majorization condition defined in their paper and $\vec \alpha$ is the vector corresponding to the canonical form of $H$ as prescribed by~\cite{dur2001entanglement}.

We translate this statement for our Hamiltonian in~\Cref{eq:ashn_zz_omega}, for which $\vec \alpha = (g/2,g/2,h/2).$
Now,~\Cref{eq:major1} can be written as the set of inequalities:
\begin{align*}
 g \tau/2 & \geq x,\\
 (2g+h) \tau/2 &\geq x+y+ z,\\
 (2g-h) \tau/2 &\geq x+y- z,
\end{align*}
while~\Cref{eq:major2} translates to
\begin{align*}
    g   \tau /2 & \geq \pi/2 -x,\\
    (2g+h) \tau/2 & \geq \pi/2-x+y-z,\\
    (2g-h) \tau/2 & \geq \pi/2-x+y+z.
\end{align*}

The \emph{optimal time} $\tau_{opt}$ is defined to be the minimum number satisfying either set of inequalities. The following theorem shows that the optimal time can always be achievable with the AshN gate scheme:


\begin{thm}[AshN gate scheme with optimal time; see~\Cref{thm:ashn} in~\Cref{app:ashn}]
For each Weyl chamber coordinate $(x,y,z)\in W$, XY coupling strength $g\in\mathbb{R}^+$ and ZZ coupling strength $h\in[-g,g]$, let $\tau_{opt}$ be the optimal gate time defined as above.
There exist $\Omega_1,\Omega_2,\delta\in\mathbb{R}$ such that
$\exp\{-i \cdot \tau_{opt}\cdot H_R(g,h;\Omega_1,\Omega_2,\delta)\}$ has interaction coefficients $(x,y,z).$
Moreover, $\Omega_1\Omega_2\delta = 0$, meaning that at least one of the three parameters is zero. 
\label{thm:ashn_main}
\end{thm}

However, the amplitude and the detuning of the microwave drive become unbounded for gates in the neighborhood of the identity if we want to achieve the optimal gate time. In general, achieving optimal time an $\varepsilon$-neighborhood of the identity gate in the Weyl chamber requires amplitude and/or detuning of magnitude $\Omega(\frac{1}{\varepsilon})$, which can be practically infeasible. Introducing AshN-ND-EXT resolves this matter, but with non-optimal gate times for gates near the identity. We will explore this more thoroughly in~\Cref{subsec:gen_2qubit}.

\subsection{Free Single-Qubit Gates}
A bonus feature of the AshN gate scheme is that we can implement a one-parameter family of single-qubit gates with no additional overhead. We can see this by directly looking at~\Cref{eq:ashn_ham} or even the more general~\Cref{eq:ashn_zz_ham}. Define 
\begin{align*}
    \bar \phi:= \frac{\phi_1+\phi_2}{2}
\end{align*}
and
\begin{align*}
    \phi' := \frac{\phi_1-\phi_2}{2}.
\end{align*}
We observe 
\begin{align*}
    \cos \phi_1 XI - \sin \phi_1 YI = (Z_{-\bar \phi} \otimes Z_{-\bar \phi}) (\cos \phi'XI - \sin \phi' YI)  (Z_{\bar \phi} \otimes Z_{ \bar\phi})
\end{align*}
and similarly
\begin{align*}
    \cos \phi_2 IX - \sin \phi_2 IY = (Z_{-\bar \phi} \otimes Z_{-\bar \phi}) (\cos \phi' IX + \sin \phi' IY) (Z_{\bar \phi} \otimes Z_{\bar \phi}).
\end{align*}
Making explicit the dependence of the Hamiltonian in~\Cref{eq:ashn_zz_ham} on $\phi_1,\phi_2$, we have
\begin{align*}
    H_R(\phi_1, \phi_2) = (Z_{-\bar \phi} \otimes Z_{-\bar \phi}) H_R(\phi',-\phi') (Z_{\bar \phi} \otimes Z_{\bar \phi}).
\end{align*}
Therefore, by simply tuning $\bar \phi$, we can get an effective additional $Z$ gate akin to virtual $Z$ gates in the single-qubit scenario~\cite{mckay2017efficient, chen2023compiling}. And just like the virtual $Z$ case, this phase shift property does not depend on the exact pulse envelope of the single-qubit drives and thus can be directly exploited in experiments.

\section{Calibration}
\label{sec:cal}


\subsection{Calibration of a Single Gate}
\label{subsec:cartan}

To fine-tune a specific two-qubit gate, we need an efficient and accurate way of characterizing the experimentally realized gate. While calibration methods like quantum process tomography (QPT) or gate set tomography (GST) can be used for characterization of non-standard gates while being agnostic of the experimental realization~\cite{lin2022let}, such protocols can often be inefficient as they both aim to provide a comprehensive characterization of the full quantum gate.

In particular, for two-qubit gates, the Weyl chamber coordinate of a gate is generally more important than the single-qubit components, as single-qubit gates can often be realized with higher fidelity than two-qubit gates. 
We here propose a way to directly and efficiently extract the interaction coefficients \emph{without having to determine the corresponding single-qubit corrections}. Specifically, we propose the following method to convert interaction coefficient estimation to phase estimation.

Given a two-qubit gate $U\in \textbf{SU}(4)$, we define its \emph{Cartan double} as
\begin{align}
    \gamma(U):=U\cdot YY\cdot U^T\cdot YY.
\end{align}
It can be verified that $\gamma(U)=(A_1\otimes A_2)\exp\{2i\vec{\eta}\cdot\vec{\Sigma}\}(A_1^\dagger \otimes A_2^\dagger)$. As such, the interaction coefficients $\vec{\eta}$ of $U$ can be estimated with phase estimation from its Cartan double $\gamma(U)$. 

To realize $\gamma(U)$ experimentally for the AshN gate schemes, note that $H_R=H_R^T$ and consequently $U=U^T$. Phase estimation on the unitary $U\cdot YY$ then determines the interaction coefficients of $U$. Alternatively, we observe that
\begin{align*}
    &-\vartheta(H_R(g,h;\Omega_1,\Omega_2,\delta)):=YY\cdot H_R^T(g,h;\Omega_1,\Omega_2,\delta)\cdot YY=H_R(g,h;-\Omega_1,-\Omega_2,-\delta),
\end{align*}
indicating that $\Theta^{-1}(U):=YY\cdot U^T\cdot YY$ can be realized again with the AshN gate scheme, with opposite sign microwave pulse amplitudes and detuning. 

\begin{figure}
    \centering
    \includegraphics[width=0.55\textwidth]{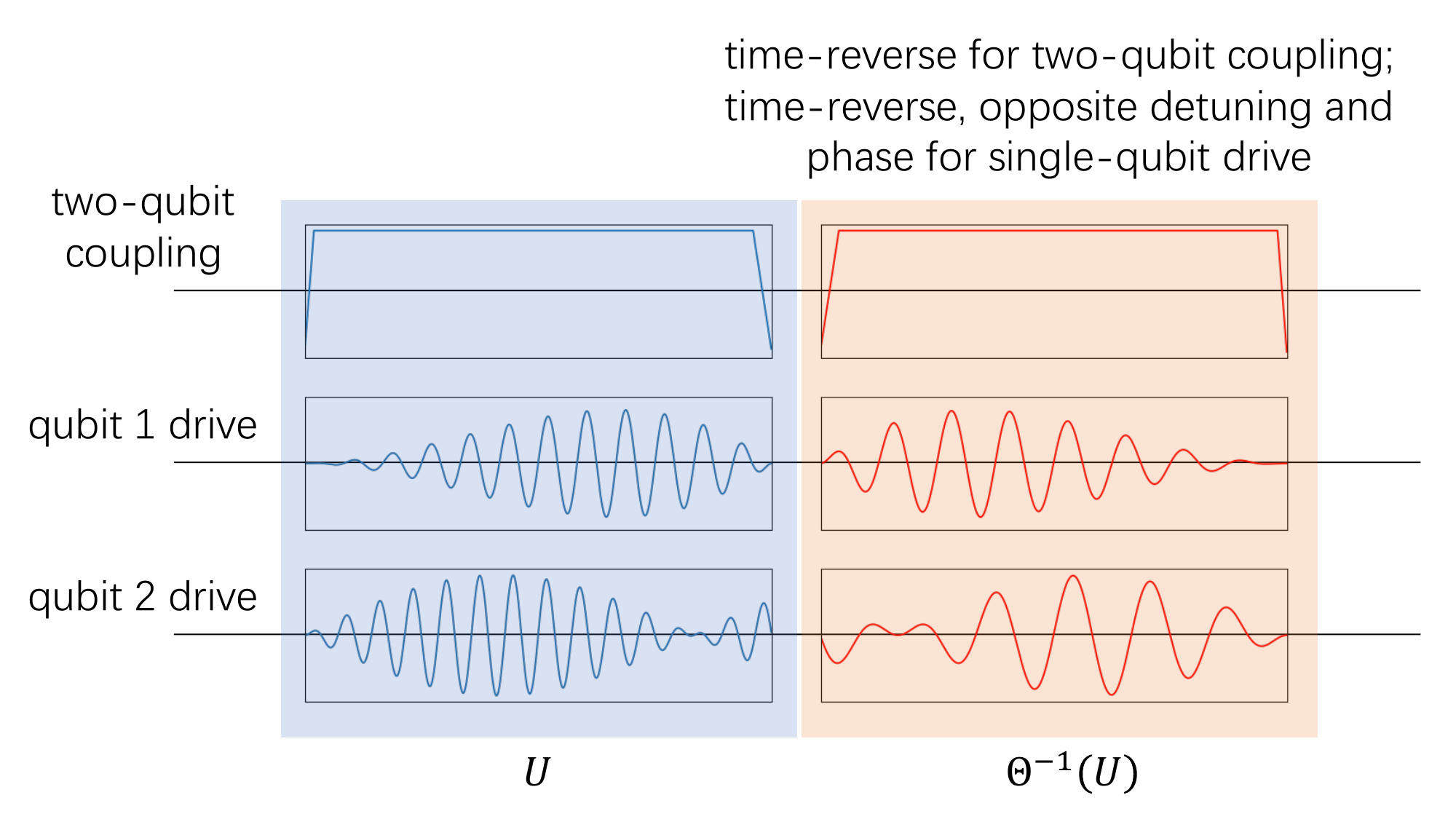}
    \caption{Illustration of the Cartan double of a given gate scheme. The Cartan double is given as $\gamma(U):=U\Theta^{-1}(U)$. To realize $\Theta^{-1}(U)$, one plays the waveform for two-qubit coupling in time reverse, and the single-qubit waveforms in time reverse, with opposite phase and opposite detuning.}
    \label{fig:cartan_double}
\end{figure}

However, it may be difficult to implement a clean time-independent Hamiltonian evolution, which we are assuming in the AshN scheme, because a perfect square pulse envelope cannot be implemented in experiments. In general, it is approximated by a trapezoidal envelope or with rise and fall amplitudes that go as $1-\cos$ or $\tanh$. Hence $H_R$ is time dependent and in general $U \neq U^T$. When time-dependent terms are present in the Hamiltonian, the time evolution unitary is given by the time-ordered integral 
\begin{align*}
    U=\mathcal{T}\exp\{-i\int_{t=0}^TdtH_R(t)\},
\end{align*}
so
\begin{align*}
    U^T=\mathcal{T}\exp\{-i\int_{t=0}^TdtH_R(T-t)\},\\ \Theta^{-1}(U)=\mathcal{T}\exp\{i\int_{t=0}^Tdt\vartheta(H_R(T-t))\}.
\end{align*}
To calibrate the interaction coefficients, we need to play the microwave pulses backwards in time interleaved with $YY$, or backwards in time with opposite phase and detuning. Assuming negligible pulse distortion, this should be achievable via a customized pulse stored on the AWG (arbitrary waveform generator).
Implementation of the Cartan double is illustrated in \Cref{fig:cartan_double}.

Note that reducing Weyl chamber coordinate calibration to phase estimation of the Cartan double is applicable to any two-qubit gate scheme, as long as we can realize $\Theta^{-1}$ or the transpose of a given gate. 

\subsection{Calibration of an Instruction Set}
The AshN gate scheme generates a continuous family of gates. While it is possible to develop calibration schemes to optimize the fidelity of one or a few specific gates, calibrating \emph{all} gates in the gate set one by one is impossible as there are infinitely many. 
However, since all AshN gates share a common gate scheme, we should be able to describe the fidelity of each gate with respect to the control parameters via a relatively simple mathematical model. This opens up the possibility for calibration without having to optimize each gate individually. 

More formally, we assume that the control scheme for each gate is given by a parameterized mathematical model. We make a distinction between the \emph{model parameters} that specify the mathematical model that maps ideal \emph{gate parameters} from Algorithm~\ref{alg:ashn} to its corresponding \emph{control parameters} that are directly used in experiments. One example of such a model describes the output single qubit drive (control parameter) as linearly proportional (ratio is model parameter) to its digital representation in the program (gate parameter). By adopting such models, we reduce the task of calibrating individual gates to determining the best-fit model parameters of the said model, thereby enabling high-fidelity realization of all gates. This could be done by taking the average fidelity of the gate set using fully randomized benchmarking (FRB~\cite{kong2021framework}) or more general universal RB schemes~\cite{PRXQuantum.3.030320} under different distributions as objective functions and performing black-box optimization to find the best-fit model parameters.

\section{Applications}
The AshN gate scheme can provide a performance boost for the quantum device in multiple different ways. Here we list some important examples.

\subsection{Arbitrary Two-Qubit Gates}\label{subsec:gen_2qubit}
The most obvious application of the AshN scheme is that we can directly implement two-qubit gates up to single-qubit gate corrections. If we can include all the gates the scheme can realize into our instruction set, we can directly implement any two-qubit circuit with just one two-qubit gate, while instruction sets with just a $\CNOT$ or $\iSWAP$ would need up to three two-qubit gates~\cite{ye2004super}. Of course, gate count is not always an accurate cost metric for a quantum circuit. As we noted above, decoherence is often one of the main sources of gate error, and so a more useful metric is the gate time. 

\begin{figure}
    \centering
    \includegraphics[width=0.65\textwidth]{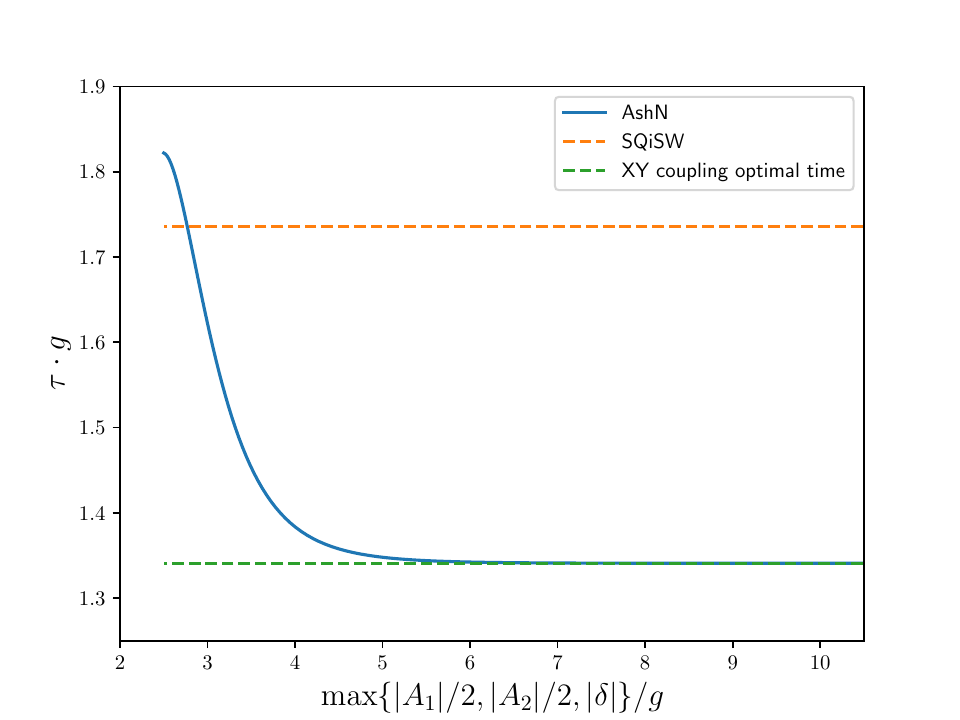}
    
    \caption{ The average two-qubit interaction time versus the required microwave drive amplitude and detuning (both normalized with respect to $g$). We assume no $ZZ$ coupling. As a comparison, we include the average two-qubit gate time for compiling Haar random two-qubit gates using SQiSW from~\cite{huang2023quantum}. Note that total gate time using SQiSW would be longer due to additional single-qubit gates between the two applications of the SQiSW gate. Two-qubit gate time with iSWAP and CZ each via flux tuning are $4.712g^{-1}$ and $6.664g^{-1}$ respectively, well beyond the range of this plot.}
    \label{fig:gate_time_amp}
\end{figure}

Recall we need to fix a cutoff $r$ in order to avoid unbounded drive amplitudes and detunings. However, this cutoff can lengthen gate times. We directly plot the trade-off between the two-qubit gate time using the AshN scheme averaged over Haar random two-qubit gates and the maximum required microwave amplitude and detuning in \Cref{fig:gate_time_amp}. We assume $h=0$ so that we can apply~\Cref{eq:r_cutoff}. When the amplitudes and detunings can be unbounded, we only use AshN-ND or AshN-EA and achieve an average gate time\footnote{For simplicity we ignore the time taken to ramp the pulse up and back down in all of our predicted gate times. Furthermore, in experiments we typically need to calibrate the exact gate time to optimize a metric such as average fidelity. } of $(\frac{7\pi}{16}-\frac{19}{180\pi})g^{-1} \approx 1.341g^{-1}$. Note again that this is the theoretically optimal two-qubit gate time we can obtain as shown in~\Cref{subsec:time_op}. In comparison, using $\SQiSW$ using flux tuning with a gate time of $\frac{\pi}{4g}$ would lead to an average two-qubit gate time of $\approx 1.736 g^{-1}$~\cite{huang2023quantum}, which is approximately 1.29 times slower. Note that we are not including single-qubit gate overheads necessary for using $\SQiSW$. A flux-tuned $\iSWAP$ gate with a gate time of $\frac{\pi}{2g}$ would entail an average two-qubit interaction time of $\frac{3\pi}{2g} \approx 4.712g^{-1}$, which is 3.51 times slower. If we use a flux-tuned $\CZ$ gate on a transmon device with a gate time of $\frac{\pi}{\sqrt{2}g}$, the average gate time would be $\frac{3\pi}{\sqrt{2}g} \approx 6.664g^{-1}$, which is astoundingly 4.97 times slower.

It is important to see whether these amplitudes and detunings are feasible in superconducting qubit experiments. For conciseness, we will only consider transmon and fluxonium devices that are capacitively coupled with device parameters commonly found in experiments. For transmon devices, coupling magnitudes can range from $5 \times 2\pi$ to $40 \times 2\pi$ MHz~\cite{krantz2019quantum}, while fluxonium devices, due to small charge matrix elements, typically have smaller couplings ranging from $5 \times 2 \pi$ to $7 \times 2 \pi$ MHz~\cite{bao2022fluxonium, moskalenko2022high, PhysRevApplied.20.024011}. Now, looking at~\Cref{fig:gate_time_amp}, if we want to be within $10\%$ of the minimum value of $1.341 g^{-1}$, we can afford a cutoff $r$ as large as $r=1.1$. In this case, $\pi/r +1/2 \approx 3.356$. This would translate to a maximum detuning or amplitude of $34 \times 2\pi$ to $260 \times 2 \pi$ MHz for transmon devices and $34 \times 2 \pi$ to $46 \times 2\pi$ MHz for fluxonium devices. Such detunings are clearly manageable. For transmon devices, such large drive amplitudes can lead to leakage and would need pulse corrections such as DRAG~\cite{motzoi2009simple, gambetta2011analytic, de2015fast}, but this is not what we assume in our Hamiltonian as per~\Cref{eq:square_pulse}. However, for couplings on the lower end, amplitudes up to $49 \times 2 \pi$ MHz are reasonable if the transmon qubits have sufficiently large anharmonicities as shown by numerical simulations in~\cite{guo2018dephasing}. As for fluxonium, single-qubit gates as fast as 10 ns can be done with negligible leakage~\cite{bao2022fluxonium}. This is comparable
to what we need given the smaller couplings for
fluxonium. Hence, by appropriately engineering the device parameters, \emph{the required amplitudes for coming reasonably close to the optimal time are feasible for superconducting devices.} Note again that this parameter engineering can be applied to simultaneously optimize all possible two-qubit gates since the underlying gate scheme is the same. Also note that this amplitude requirement is for when we want to generate the entire polytope contained in the Weyl chamber corresponding to the cutoff chosen. For specific gate classes, the required amplitude can be considerably smaller. For example, $\iSWAP$ is inside the $r=\pi/2$ polytope, which requires a maximum amplitude or detuning of $2.5g$. However, $\iSWAP$ can be realized purely with flux tuning and does not even need a microwave drive.

\subsection{Arbitrary \texorpdfstring{$n$}{n}-Qubit Gates}
\label{subsec:su_n}
\begin{figure*}
    \begin{minipage}{0.60\textwidth}
    \includegraphics[width=1.00\textwidth]{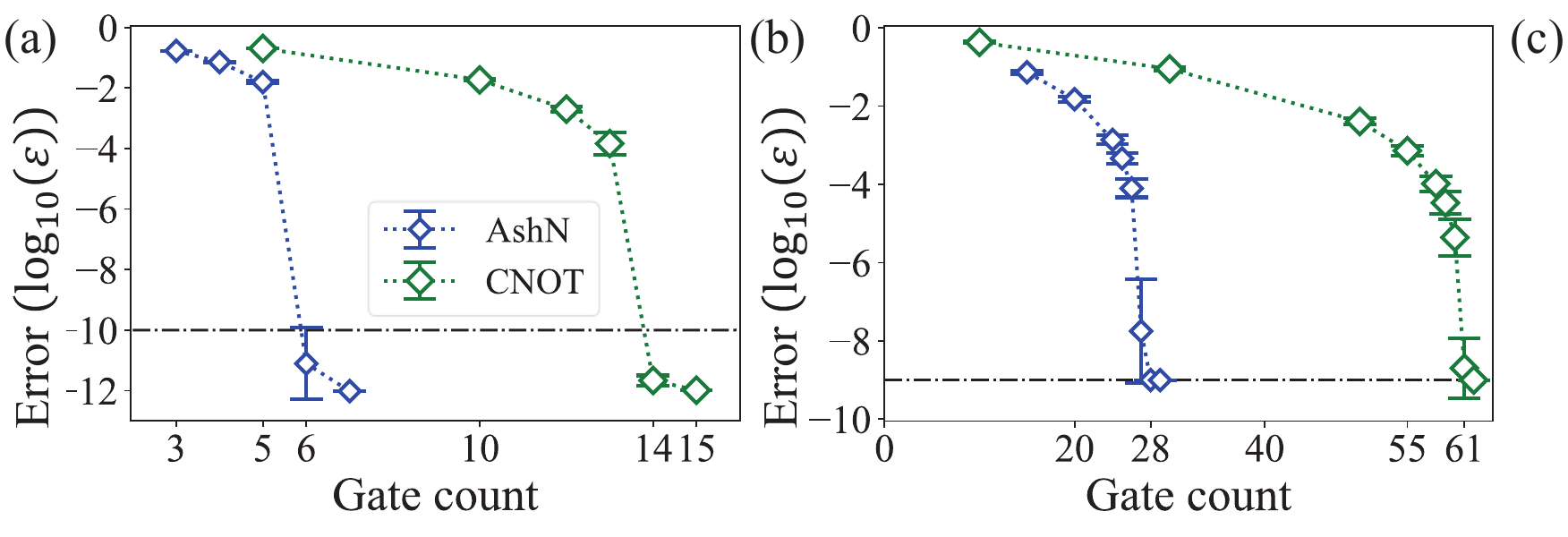}
    \end{minipage}
    \begin{minipage}{0.44\textwidth}
    \begin{tabular}{|c|c|c|c|}
    \hline
     & $3$-qubit & $4$-qubit & $n$-qubit \\
    \hline
    CNOT (N)~[*] & $14$ & $61$ & N/A \\
    \hline
    AshN (N)~[*] & $6$ & $27$ & N/A \\
    \hline
    CNOT (A)~\cite{shende2005synthesis} & $20$ & $100$ & $\sim\frac{23}{48}4^n$\\
    \hline
    AshN (A)~[*] & $11$ & $68$ & $\sim\frac{23}{64}4^n$\\
    \hline
    \end{tabular}
    \end{minipage}
    \caption{(a) We plot the logarithmic decomposition error $\log_{10}(\epsilon)$ as a function of the two-qubit gate count, where the two-qubit gate count realizable via the AshN scheme is plotted with blue dots and the $\CNOT$ gate count is plotted with green. (b) The same as (a), but for $n=4$. (c) Comparisons of the numerical (N) and analytical (A) state-of-the-art gate counts for gate decomposition using arbitrary two-qubit gates versus $\CNOT$ gates. The results provided by our paper are marked by [*]. All numerical results match corresponding lower bounds. Note that the 4-qubit AshN analytic result is a special case of the $n$-qubit result.}
    \label{fig:gate_syn}
\end{figure*}

Quantum gate decomposition, also known as quantum circuit synthesis~\cite{barenco1995elementary}, splits arbitrary $n$-qubit unitaries into a sequence of single- and two-qubit gates that are local and more amenable to experimental realization. This is an essential task for implementing computational tasks on quantum devices. Here, we show through extensive numerical evidence as well as analytical proofs that the AshN gate scheme is able to provide a significant reduction in gate count for different scenarios.

The vast majority of studies on quantum gate decomposition use $\CNOT$ as the native two-qubit gate. By counting parameters, it was shown that at least $\lceil\frac{1}{4}(4^n-3n-1)\rceil$ $\CNOT$ gates are required for an arbitrary $n$-qubit gate, assuming arbitrary single-qubit gates are available~\cite{shende2004smaller,shende2004minimal}. The conventional QR decomposition yields a sequence of elementary gates containing $O(n^34^n)$ two-qubit gates~\cite{barenco1995elementary,cybenko2001reducing}. An improved approach via the Gray code basis provides the upper bound of $4^n-2^{n+1}$ for circuit synthesis using $\CNOT$ gates~\cite{aho2003design,knill1995approximation,mottonen2004quantum,vartiainen2004efficient}. The strategy with the lowest known asymptotic $\CNOT$ count is the quantum Shannon decomposition (QSD)~\cite{shende2005synthesis}, which gives a gate count of $\frac{23}{48}4^n-\frac32 2^n+\frac{4}{3}$. This is approximately twice the theoretical lower bound. 

Now, the AshN scheme can directly implement any two-qubit gate up to single-qubit corrections. Since single-qubit gates are usually cheap, this motivates the problem of quantum gate decomposition \emph{using arbitrary two-qubit gates}, that is, the special unitary group $\textbf{SU}(4)$. For this problem, a theoretical lower bound of $\lceil\frac{1}{9}(4^n-3n-1)\rceil$ two-qubit gates is known~\cite{yu2013optimal}. In this paper, we establish the following gate count results for quantum gate decomposition using arbitrary two-qubit gates from $\textbf{SU}(4)$. We prove this in~\Cref{app:decomp}.

\begin{thm}\label{thm:GateSyn}
Any three-qubit gate can be decomposed into a sequence of single-qubit gates and at most $11$ generic two-qubit gates. The decomposition can be made fully explicit. In addition, an arbitrary $n$-qubit unitary can be implemented in a circuit containing no more than $\frac{23}{64}4^n-\frac{3}{2}2^n$ generic two-qubit gates.
\end{thm}

\Cref{thm:GateSyn} gives a two-qubit gate count that is $3/4$ of that of the number of $\CNOT$ gates required for $n$-qubit unitaries via QSD. For the case of $n=3$, an analytic proof achieves $20$ $\CNOT$ gates~\cite{rakyta2022approaching}, while the lower bound is $14$ $\CNOT$ gates. Thus, for three-qubit gates~\Cref{thm:GateSyn} gives a two-qubit gate count slightly over a half of what is analytically provable for $\CNOT$. The number of two-qubit gates required lies within a factor of $2$ from the theoretical lower bound of $6$ for three-qubit gates and a factor of $3$ for arbitrary $n$-qubit unitaries. This indicates \emph{a significant potential resource reduction for using the AshN gate scheme instead of $\CNOT$ as is usually considered in the literature.}

On top of~\Cref{thm:GateSyn}'s analytic results showing gate count reductions using AshN compared to $\CNOT$, we also perform numerical experiments to explore the number $N$ of arbitrary two-qubit gates or $\CNOT$ gates required for gate decomposition at $n=3$ and $n=4$. We label the qubits from $0$ to $2$ ($3$) for three (four)-qubit gate decomposition. For gate decomposition using arbitrary two-qubit gates, we employ a circuit composed of two-qubit gates sequentially acting on qubit pairs $(0,1)$, $(0,2)$ (and $(0,3)$ for $n=4$) to approximate the target unitary.\footnote{Note that for $n=3$ this is the most general circuit structure possible since we can always add appropriate $\SWAP$ gates.} For gate decomposition using $\CNOT$ gates, we consider a similar circuit by replacing every arbitrary two-qubit gate with a $\CNOT$ gate and adding single-qubit rotations between neighboring $\CNOT$ gates. For each gate count $N$, we Haar randomly generate $1000$ target unitaries and use the QFactor package~\cite{kukliansky2023qfactor} to optimize the parameters of the approximating circuit. The decomposition error is measured using the distance 
\begin{align*}    
\text{dist}(U,V)=1-\frac{\abs{\tr[U^\dagger V]}}{2^n}
\end{align*}
between the target unitary $U$ and approximation circuit $V$. For each target unitary, the optimization halts if the decomposition error is smaller than a threshold, which is set to $10^{-10}$ and $10^{-9}$ at $n=3$ and $n=4$, respectively. We finally compute the average decomposition error over all $1000$ target unitaries. In \Cref{fig:gate_syn}(a), we compare the number of $\CNOT$ and arbitrary two-qubit gates required for $n=3$. We observe that the two-qubit gate count decreases by more than $50\%$ if we consider arbitrary two-qubit gates instead of $\CNOT$ to achieve the same decomposition error. We also observe a sharp reduction in $\epsilon$ when the number of the $\CNOT$ gates and generic two-qubit gates exceeds the theoretical lower bound $14$ and $6$, respectively, for $n=3$ from dimension counting which indicates that the lower bounds might be tight. We remark that our numerical experiment indicates that $14$ $\CNOT$ gates might be enough for this task instead of $15$ as widely believed in literature~\cite{rakyta2022approaching,Madden2022Best}. We achieve this through the advanced optimization algorithm provided by the QFactor package. 
We also consider gate decomposition for $n=4$ and see a similar gain from using AshN in~\Cref{fig:gate_syn}(b). We summarize the numerical and state-of-the-art analytical results for gate decomposition in Table~\ref{fig:gate_syn}(c). 
Note that if we assume the numerical results are true, in particular, that we can implement any three-qubit gate using $6$ arbitrary two-qubit gates, we can further deduce that an arbitrary $n$-qubit gate can be decomposed into at most $\frac{9}{32}4^n$ two-qubit gates by using the same techniques as~\Cref{thm:GateSyn}. 

\subsection{Quantum Volume}
We know that gate count is not the deciding factor for the fidelity of a quantum circuit. We also need to take into account the errors of each gate. Quantum volume~\cite{cross2019validating} is a metric used to assess the overall capability of a quantum computer, taking into account factors such as the number of qubits, error rates, expressibility of native gate sets, quality of compilers, and qubit connectivity.
We conduct numerical experiments that compare quantum volume using $\CNOT$, $\SQiSW$, and our AshN scheme under different noise levels, ceteris paribus.

More explicitly, quantum volume measures the largest square random quantum circuit that a quantum computer can successfully execute. 
Here a square random circuit of size $d$ consists of $d$ qubits and $d$ layers of gates.
In each layer, $d$ qubits are randomly divided into $d/2$ pairs (here we assume $d$ is even for simplicity), and a Haar-random two-qubit unitary is applied to each pair. This circuit defines a unitary $U\in \textbf{SU}(2^d)$. The ideal output distribution is $p_U(x)=\abs{\bra{x}U\ket{0^d}}^2$.
The heavy outputs are $$H_U=\{x\in \{0, 1\}^d|P_U(x)>p_{\text{med}}\},$$
where $p_{\text{med}}$ is the median of $\{p_U(x)\}_{x\in \{0, 1\}^d}$.
Now we implement (or numerically simulate) $U$ on a quantum device and compute the probability of heavy outputs.
We say the quantum device passes the random circuit $U$ if the heavy output proportion is at least $2/3$. 
The quantum volume of the device is defined as $2^d$, where $d$ is the largest size for which the device passes all size-$d$ square random circuits.

In our numerical experiments, the quantum device has a 2D grid connectivity graph. All two-qubit gates, including the $\SWAP$ gates used for qubit routing, are compiled using using either the flux-tuned $\CZ$~\cite{strauch2003quantum,dicarlo2009demonstration}, flux-tuned $\SQiSW$~\cite{arute2020observation, google2020hartree,bialczak2010quantum,mi2021information, abrams2020implementation, PRXQuantum.2.040309, goerz2017charting,huang2023quantum}, or the AshN gate scheme with $h=0$.
We assume that every native gate is subject to depolarizing noise.
Single-qubit gates have a fixed error rate of $0.1\%$.
The error rate of flux-tuned $\CZ$ gates, $e_{\CZ}$, ranges from $0.7\%$ to $1.7\%$.
The error rate of other two-qubit gates can be determined by the error rate of $\CZ$ assuming that the error rate is proportional to the gate time. Explicitly, the gate time of $\CZ$ is taken to be $\frac{\pi}{\sqrt{2}g}$, the gate time of $\SQiSW$ is $\frac{\pi}{4g}$, and the gate time of AshN is given by Algorithm~\ref{alg:ashn} where we take $r=0$ (optimal time) and $r=1.1$ (physically feasible as discussed in~\Cref{subsec:gen_2qubit}).

\begin{figure}[ht!]
    \centering
    \includegraphics[width=0.6\linewidth]{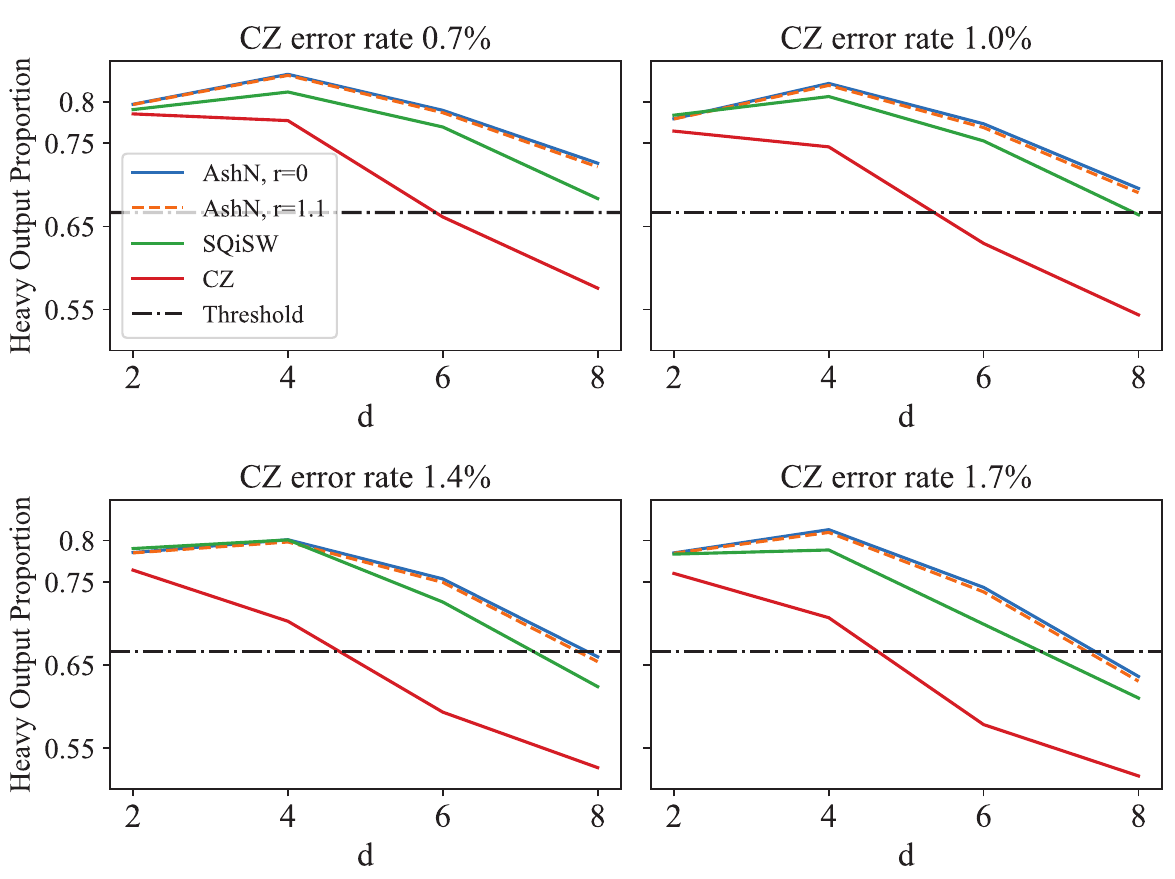}
    \caption{Heavy output proportion as a function of the size $d$ for different depolarizing noise rates. We assume here a 2D grid connectivity graph. Note that the blue solid line and the orange dotted line almost coincide, meaning that choosing a practical cutoff $r=1.1$ does not appreciably compromise performance.} 
    \label{fig:quantum_volume}
\end{figure}

We conduct the numerical experiments using the quantum volume module in Cirq~\cite{cirq_developers_2023_8161252}.
The results are shown in~\Cref{fig:quantum_volume}, where the heavy output proportion is approximated by averaging over 1350 samples of square random circuits.
For each error rate, the AshN gate scheme achieves a higher heavy output proportion, thus outperforming $\SQiSW$ and $\CNOT$ with respect to quantum volume.
Moreover, our results show that $r=1.1$ achieves almost the same performance as the no cutoff case $r=0$.
Thus, \emph{superconducting quantum devices using the AshN gate scheme can have superior overall computational power.}

\subsection{Implementing Special Gate Classes}
The AshN scheme not only provides access to all two-qubit gates up to local equivalence, it can also provide advantages over traditional schemes to implement special local equivalence classes of gates. We will consider special gate classes notable for their application in quantum information processing. We explicitly give the gate parameters for each special gate class in~\Cref{tab:special_gates} for $h=0$ (For nonzero $h$, the parameters can be computed from~\Cref{alg:ashn} with $r=0$.). Note that we use the parameters $\tau, A_i, 2\delta$ that are more standard in experiments: corresponding to gate time, drive amplitude, and drive detuning, respectively. The decimal numbers are to 4 significant figures. 
\begin{table}[htp]
    \centering
    \begin{tabular}{|c|c|c|c|c|}
         \hline
         Gate Class & $\tau$ & $A_1$ & $A_2$ & $2\delta$ \\
         \hline
         [CNOT] & $\frac{\pi}{2g}$ & $- \sqrt{15} g$ & 0 & 0 \\
         \hline
         [SWAP] & $\frac{3\pi}{4g}$ & $-2.108g$ & $2.108g$ & $-1.528g$ \\
         \hline
         [B] & $\frac{\pi}{2g}$ & $-2.238g$ & 0 & 0 \\
         \hline
    \end{tabular}
    \caption{Gate parameters for special gate classes assuming $h=0$. }
    \label{tab:special_gates}
\end{table}
\subsubsection*{Fault-Tolerant Quantum Computation}
The $\CNOT$ gate local equivalence class $(\pi/4,0,0)$ is the most widely used two-qubit gate family in fault-tolerant quantum computation. 
The gate parameters are shown in~\Cref{tab:special_gates}. 

An alternative way to achieve a $\CNOT$ equivalent gate on flux-tunable devices would be to use two $\SQiSW$ gates as proposed in~\cite{moskalenko2021tunable, moskalenko2022high, kubica2022erasure}, which achieves the same gate time of $\frac{\pi}{2g}$ but incurs single-qubit gate overheads. This can be a problem since single-qubit gate times are not always negligible compared to two-qubit gate times. And just like these proposals, we can also address unwanted $ZZ$ coupling, but this time by simply by making an appropriate modification of the scheme parameters. In fact, for the $\CNOT$ gate class we can give a simple form for the gate parameters in the presence of $ZZ$ coupling:
\begin{align*}
    & T = \frac{\pi}{2g},\, A_1 = -  \frac{\sqrt{16g^2-(g-h)^2} + \sqrt{16g^2-(g+h)^2}}{2},\\
    & A_2 = -  \frac{\sqrt{16g^2-(g-h)^2} - \sqrt{16g^2-(g+h)^2}}{2},\, 2\delta=0.
\end{align*}
It is worth computing which exact $\CNOT$ class gate we obtain:
\begin{align*}
    U(T;g,\Omega_1,\Omega_2,\delta) \vert_{p_{[\CNOT]}} = 
    \frac{1}{\sqrt{2}}
    \begin{bmatrix}
    1&0&0&-i\\
    0&1&-i&0\\
    0&-i&1&0\\
    -i&0&0&1
    \end{bmatrix}.
\end{align*}
Upon inspection, this is the $XX_{\frac \pi 2}$ rotation, also known as the M\o lmer-S\o rensen gate in trapped-ion literature~\cite{sorensen2000entanglement}. Conveniently, there are elegant circuits that use this gate to compile the syndrome measurement circuits for the surface code~\cite{schwerdt2022comparing}. Note that the gate scheme in~\cite{guo2018dephasing} realizes this gate with a similar physical implementation, but their scheme takes the limit of large drive amplitudes, which could exacerbate leakage. The AshN gate scheme applied to the $\CNOT$ class was first considered in~\cite{Ding_Wu_Zhao}, and the observation that $ZZ$ coupling can be accommodated was first made in~\cite{huihai_private}. 

Another way to achieve $[\CNOT]$ is to use two iSWAP gates interleaved with single-qubit gates. But this doubles the gate time compared to using $\SQiSW$. 
Yet another possibility is to directly implement a $\CZ$ gate by tuning the $\vert 11\rangle$ state to a higher level state such as $\vert 20 \rangle$~\cite{strauch2003quantum,dicarlo2009demonstration}. However, the gate time for this scheme is $\frac{\pi}{\sqrt{2}g}$, which is longer than the $\SQiSW$ method by a factor of $\sqrt{2}$. 

\subsubsection*{Qubit Routing}
One of the most significant overheads for implementing NISQ algorithms on current devices is qubit routing. There is therefore a strong emphasis on implementing the $\SWAP$ gate on physical devices. Using AshN-EA, we can directly implement a $\SWAP$ gate. 

The scheme parameters to realize a $\SWAP$ gate are given in~\Cref{tab:special_gates}.
The careful reader would notice that our scheme can only guarantee a gate in the SWAP local equivalence class, but remarkably if we plug in these exact parameters to the Hamiltonian evolution, we obtain the gate
\begin{align*}
    U(T;g,\Omega_1,\Omega_2,\delta)\vert_{p_{[\SWAP]}} = ZZ\cdot \SWAP.
\end{align*}
Given that there are already single-qubit phase corrections that we need to make because the AshN scheme involves tuning the frequency of the qubits, we can simply add this $Z \otimes Z$ term so that there is no additional single-qubit gate overhead than what is already necessary.

When there is $ZZ$ coupling, we can attain the optimal time with respect to the modified Hamiltonian with the $ZZ$ coupling using the corresponding modified AshN scheme. Interestingly, it is straightforward to show that this new optimal time is shorter than that without $ZZ$ coupling:
\begin{align*}
    \tau_\text{opt} = \frac{3 \pi}{4(g+\vert h \vert/2)} \leq \frac{3 \pi}{4g}.
\end{align*}
That is, $ZZ$ coupling can actually improve $\SWAP$ fidelity.

Alternatives to realizing $[\SWAP]$ on the same architecture include compiling it using multiple applications of a two-qubit gate, such as three $\CZ$'s, three $\iSWAP$'s, or three $\SQiSW$'s~\cite{shende2003recognizing, huang2023quantum}. The latter approaches can be costly in terms of overall gate time and may involve additional overhead in single-qubit gates. There are also other known schemes to directly implement $\SWAP$, including~\cite{foxen2020demonstrating,nguyen2022programmable} which provide gate schemes to realize all excitation number conserving gates. That is, $U$ preserves the subspaces $\Span\{\vert 00\rangle\}$, $\Span\{\vert 01\rangle, \vert 10\rangle\}$, $\Span\{\vert 11\rangle\}$. The scheme in~\cite{foxen2020demonstrating} is roughly equivalent to an $\iSWAP$ concatenated with a $\CZ$, for which the total two-qubit gate time is $\frac{(\sqrt{2}+1)\pi}{2g}$, which is longer than that of the AshN scheme by a factor of $4(\sqrt{2}+1)/3 \approx 3.219$. The scheme in~\cite{nguyen2022programmable} involves Floquet qubits, a very different platform, and is therefore not directly comparable with our scheme.

\subsubsection*{Unlocking the B Gate}
The B gate~\cite{zhang2004minimum} local equivalence class is defined as the class of gates with interaction coefficients $(\pi/4,\pi/8,0)$. It is the only Weyl chamber coordinate where two applications of the gate, interleaved with single-qubit gates, generate the whole Weyl chamber.\footnote{This property is actually unique to the $B$ gate class. For any two-qubit gate $U$ where two applications of $U$ generate both the identity and the SWAP gates, $U,U^\dag$ and $\SWAP \cdot U^\dag$ must share the same Weyl chamber coordinate. This uniquely determines the necessary Weyl chamber coordinate of $(\frac\pi4,\frac\pi8,0)$.} Using the AshN gate scheme (more specifically, AshN-ND), we find that a gate locally equivalent to the B gate can be realized using the parameters given in~\Cref{tab:special_gates}.
We note that an experimental proposal to implement a variety of gates on superconducting devices, including the B gate, was given in~\cite{niskanen2006tunable} involving a tunable coupling scheme between flux qubits. The tunable coupling scheme was experimentally realized in~\cite{harrabi2009engineered}, but the B gate was not mentioned. A theoretical proposal for the $B$ gate and the related $\sqrt{B}$ gate for ion trap platforms is discussed in~\cite{selvan2023mirror}. While this paper was undergoing peer review, a paper realizing the $B$ gate with two pulses calibrated together was posted to the arXiv~\cite{wei2023native}.

\section{Discussion}\label{sec:dis}
We establish a gate scheme that combines mature techniques for quantum gate implementation in superconducting devices, namely tuning qubits into resonance and applying sinusoidal microwave drives, to effectively obtain a quantum instruction set that includes all two-qubit gates up to local equivalence. Furthermore, we show that we can obtain every point in the Weyl chamber in optimal time.

For future work, the first task is to conduct an experimental demonstration of the AshN scheme. For transmon devices, leakage is a primary concern and appropriate engineering of device parameters would be needed such that anharmonicities are large and the coupling is relatively small. The small couplings could lead to slower gates, thus compromising fidelity. However, the gain from using AshN is significant enough to possibly offset this issue, with a $\sqrt{2}\times$ speed up for $\CNOT$ class gates and up to a $3.219\times $ speed up for $\SWAP$ class gates compared to other methods on flux-tunable architectures. Fluxonium devices in comparison will not suffer as much from leakage but there will still be phase errors from higher energy levels~\cite{bao2022fluxonium} that will need to be accounted for. We expect that such phase errors can be addressed via a corresponding control parameter compensation. In the end, the quality of the calibrated gates will have to be demonstrated in experiments. The entire two-qubit gate set can be characterized using techniques such as fully randomized benchmarking, while individual gates can be characterized using its interleaved version~\cite{huang2023quantum, kong2021framework}. Interesting experiments to run using AshN also include simulating the Heisenberg XYZ model, analogous to simulating the XXZ model as mentioned in~\cite{nguyen2022programmable}, and implementing higher-qubit unitaries such as three-qubit gates as we analyzed in~\Cref{subsec:su_n}.

Now, the need to calibrate each instruction pulse before computation is a crucial consideration when expanding the quantum instruction set. While some instruction set evaluation methodologies such as \cite{LMM+21} have discussed this at length, it still requires further exploration and experimental validation. For instance, interpolation of calibrated parameters is a possible but unproven method. If proven effective, this approach could significantly reduce calibration overhead. Furthermore, recent schemes like cross entropy benchmarking (XEB)~\cite{BIS+18} and fully randomized benchmarking (FRB)~\cite{huang2023quantum} have been proposed to characterize a generic two-qubit gate. These schemes enable a \textit{reconfigurable} quantum instruction set, where calibration of instruction pulses is performed before computation. 


Lastly, we have presented a gate scheme that spans all of $\textbf{SU}(4)$ modulo single-qubit gates, leading to highly optimized quantum code. However, achieving optimal instruction count in practice requires developing practical compilation schemes that can effectively leverage $\textbf{SU}(4)$ as an instruction set in various benchmarks and computational tasks. There are numerous avenues to explore. A specific conjecture is that $6$ generic two-qubit gates are sufficient to compile an arbitrary three-qubit gate. The techniques developed for this problem may potentially generalize to compiling an $n$-qubit unitary using $(n-1)$-qubit unitaries. These conjectures give only a glimpse of the new and fascinating problems in quantum compilation inspired by the rich physics of quantum computing hardware. 

\section{Acknowledgements}
We would like to thank Juan Ignacio Cirac, Xiaotong Ni, Yaoyun Shi, Ziang Wang, Feng Wu, Fei Yan, Jun Zhang, and Hui-Hai Zhao for helpful discussions. This work was supported by Alibaba Group through the Alibaba Research Intern Program and conducted when QY was a research intern at Alibaba Group. DD would like to thank God for all of His provisions.

\appendix

\section{Proof of the AshN gate scheme}
\label{app:ashn}

\subsection{Problem Statement \& Canonicalization}
\label{sec:ps}
We study the unitary
\begin{align*}
    U(\tau;g,h;\Omega_1,\Omega_2,\delta):=\exp\{-i\cdot H_R(g,h;\Omega_1,\Omega_2,\delta)\cdot \tau\},
\end{align*}
where 
\begin{align*}
    &H_R(g,h;\Omega_1,\Omega_2,\delta):=\frac g 2 (XX + YY) + \frac h 2 ZZ +\Omega_1(XI+IX)  +\Omega_2 (XI-IX) + \delta (ZI + IZ)
\end{align*}
and $|h|\leq g$. For sake of simplicity we normalize the Hamiltonian by setting $g=1$ and $h\in[-1,1]$, and let 
\begin{align*}
    &H_R(h;\Omega_1,\Omega_2,\delta):=\frac 1 2 (XX + YY + h ZZ) + \Omega_1(XI+IX)  +\Omega_2 (XI-IX) + \delta (ZI + IZ)
\end{align*}

\subsubsection{Problem Statement}
We study the Weyl chamber coordinate associated with the unitary $U$, and show that every Weyl chamber coordinate can be achieved within its corresponding optimal time. Specifically,
for each $h$ and interaction time $\tau$, the minimum time to reach the Weyl chamber coordinates $(x,y,z)$, where $\pi/4\geq x\geq y\geq |z|$, via alternately evolving under $H$ and arbitrary single-qubit Hamiltonians $H_0\otimes H_1$ is given by $\tau_{opt}:=\min\{\tau_1,\tau_2\}$ where $\tau_1$ satisfies~\cite{hammerer2002characterization} 
\begin{align*}
    \tau_1 &\geq 2x,\\
    (2+h)\tau_1&\geq 2(x+y+z),\\
    (2-h)\tau_1&\geq 2(x+y-z).
\end{align*}
and $\tau_2$ satisfies
\begin{align*}
    \tau_2 &\geq \pi-2x,\\
    (2+h)\tau_2&\geq 2(\frac \pi 2-x+y-z),\\
    (2-h)\tau_2&\geq 2(\frac \pi 2 - x+y+z).
\end{align*}
It can be verified that the whole Weyl chamber can be spanned in time $\pi$ for any $h\in[-1,1]$. We now state the main theorem:
\begin{thm}[AshN gate scheme with optimal time]
For each Weyl chamber coordinate $(x,y,z)\in W$ and ZZ coupling strength $h\in[-1,1]$, let $\tau_{opt}$ be the optimal gate time defined as
$$\tau_{opt}(h;x,y,z):=\min\left\{\max\{2x, \frac{2(x+y\pm z)}{2\pm h}\}, \max\{2x, \frac{2(\pi/2-x+y\mp z)}{2\pm h}\},\right\}.$$
There exist $\Omega_1,\Omega_2,\delta\in\mathbb{R}$ such that
$$ \exp\{-i \cdot \tau_{opt}(h;x,y,z)\cdot H_R(0;\Omega_1,\Omega_2,\delta)\}$$ has interaction coefficients $(x,y,z).$
Moreover, $\Omega_1\Omega_2\delta = 0$, meaning that at least one of the three parameters is zero. 
\label{thm:ashn}
\end{thm}
\noindent Since the required amplitudes in~\Cref{thm:ashn} can be unbounded, we also prove a theorem with bounded amplitudes but where time optimality can be compromised.
\begin{thm}[Bounded amplitudes AshN gate scheme]
For each Weyl chamber coordinate $(x,y,z)\in W$ and every $h\in[-1,1]$, there exist $\tau\leq \pi,$ $\Omega_1,\Omega_2,\delta$ such that 
$$ \exp\{-i \cdot \tau\cdot H_R(h;\Omega_1,\Omega_2,\delta)\}$$ has interaction coefficients $(x,y,z).$
Moreover, the drive strengths are bounded; we have $\max\{|\Omega_1+\Omega_2|,|\Omega_1-\Omega_2|,|\delta|\}\leq  \frac{2(1+|h|)}{1-|h|}+\frac{1}{2}$ and $\Omega_1\Omega_2\delta = 0$, meaning that at least one of the three parameters is zero. 
\label{thm:ashn-uniform}
\end{thm}

When $h=0$ we obtain a more detailed analysis of the tradeoff between the gate time and the drive strengths.

\begin{thm}[AshN gate scheme with zero ZZ coupling]
For each cutoff threshold $r\in[0,\pi/2]$, for each Weyl chamber coordinate $(x,y,z)\in W$, there exist $\tau(x,y,z;r)\leq \pi,$ $\Omega_1,\Omega_2,\delta$ such that 
$$ \exp\{-i \cdot \tau(x,y,z;r)\cdot H_R(h;\Omega_1,\Omega_2,\delta)\}$$ has interaction coefficients $(x,y,z).$
Moreover, the drive strengths are bounded; we have $\max\{|\Omega_1+\Omega_2|,|\Omega_1-\Omega_2|,|\delta|\}\leq  \frac{\pi}{r}+\frac{1}2$ and $\Omega_1\Omega_2\delta = 0$, meaning that at least one of the three parameters is zero. When $r=0$ we have $\tau(x,y,z;r)=\tau_{opt}(0;x,y,z)=\max\{2x,x+y+|z|\}$.
\label{thm:ashn-r-cutoff}
\end{thm}

\subsubsection{AshN subschemes}
We can prove these theorems by explicitly providing gate compilation schemes mapping Weyl chamber coordinates to the drive parameters. Specifically, we divide the entire Weyl chamber into subsets according to how the optimal gate time is achieved, and present the compilation subschemes case by case.

Before doing so, we recall the definition of the Weyl chamber, which is the set of equivalent classes $\mathbb{R}^3/\sim$ where $\sim$ is defined by:
\begin{itemize}
    \item $(x,y,z)\sim (y,x,z)\sim(z,y,x)$;
    \item $(x,y,z)\sim (x,-y,-z)$;
    \item $(x,y,z)\sim (\pi/2 + x, y,z)$.
\end{itemize}

This set of equivalence classes is conveniently identified with a subset of $\mathbb{R}^3$ as $W=\{(x,y,z)| \pi/4\geq x\geq y\geq |z|, x=\pi/4\Rightarrow z\geq 0\}$. Coordinates lying in $W$ are called \emph{canonical} and those outside of $W$ are non-canonical. In the rest of the proof, we sometimes define sets of Weyl chamber coordinates as subsets of $\mathbb{R}^3$, and the Weyl chamber coordinates involved can sometimes be non-canonical. It is to be understood that membership of a coordinate in a set or identification of two coordinates is defined under canonicalization.

We define the following set of polygons in $\mathbb{R}^3$:
\begin{align*}
    ND(h;\tau)&:=\left\{(x,y,z)|x=\frac\tau2, y\pm z\in\left[0, \min\left\{\frac{(1\pm h)\tau}2,\pi -\frac{(1\pm h)\tau}2 \right\}\right]\right\}\\
    EA_+(h;\tau)&:=\left\{(x,y,z)|x+y+z = \frac{(2+h)\tau}2,  x\geq y\geq|z|, y+z\geq (1+h)x \right\},\\
    EA_-(h;\tau)&:=\left\{(x,y,z)|x+y-z = \frac{(2-h)\tau}2,  x\geq y\geq|z|, (y-z)\geq (1-h)x \right\}.
\end{align*}

For each Weyl chamber coordinate $(x,y,z)\in W$, we define $\tau^*(x,y,z):=\arg\min_{\tau} (x,y,z) \in A_\tau$, where
$$A_\tau:=ND(h;\tau)\cup EA_+(h;\tau) \cup EA_-(h;\tau),\tau\in[0,\pi].$$

We symbolically verified that $\tau^*$ is well defined on $W$, and that $\tau^* =\tau_{opt}\leq \pi$\footnote{The verification is done in Mathematica and is left out from the proof for sake of clarity.}. Therefore, it suffices to prove that every point in $A_\tau$ is achievable under time-invariant Hamiltonian evolutions within time $\tau\leq \pi$, provided that appropriate single-qubit Hamiltonians are chosen. Specifically, we have the following lemmas. 
\begin{lem}[AshN-ND]
    For any $h\in[-1,1]$, $\tau \in(0, \pi]$, and Weyl chamber coordinates $(x,y,z)\in ND(h;\tau)$,
    there exist $\Omega_1,\Omega_2$ such that the interaction coefficients for $U(\tau;h;\Omega_1,\Omega_2,0)$ are $(x,y,z)$ after canonicalization. Moreover, $\Omega_1,\Omega_2$ can be taken such that  $|\Omega_1|,|\Omega_2|\leq \frac{\pi}{2\tau}$.
    \label{lem:nd}
\end{lem}



\begin{lem}[AshN-EA+]
    For any  $h \in[-1,1]$, any $\tau \in(0, \pi]$, for any Weyl chamber coordinates $(x,y,z)\in EA_+(h;\tau)$, there exist $\Omega_1,\delta$ such that the interaction coefficients for $U(\tau;h;\Omega_1,0,\delta)$ are $(x,y,z)$. Moreover, $\Omega_1,\delta$ can be taken such that $\vert\Omega_1\vert,\vert\delta\vert\leq (1-h)\frac \pi \tau + \frac{1}{2}$.
    \label{lem:ea}
\end{lem}
\noindent The gate scheme for the AshN-EA- sector can be derived from AshN-EA+ according to the following corollary.
\begin{cor}[AshN-EA-]
    For any  $h \in[-1,1]$, any $\tau \in(0, \pi]$, for any Weyl chamber coordinates $(x,y,z)\in EA_-(h;\tau)$, there exist $\Omega_2,\delta$ such that the interaction coefficients for $U(\tau;h;0,\Omega_2,\delta)$ are $(x,y,z)$. Moreover, $\Omega_1,\delta$ can be taken such that $|\Omega_1|,|\delta|\leq (1+h)\frac \pi \tau + \frac{1}{2}$.
    \label{cor:ea}
\end{cor}

\begin{proof}
    For every $h\in[-1,1]$ and $(x,y,z)\in EA_-(h;\tau)$, we want to prove that there exist $\Omega,\delta\in[-((1+h)\pi/\tau + \frac{1}{2}), (1+h)\pi/\tau + \frac{1}{2}]$ such that the interaction coefficients for $U(\tau;h;0,\Omega_2,\delta)$ are $(x,y,z)$. We consider a dual problem of generating the Weyl chamber coordinates $(x,y,-z)$ with single-qubit time-invariant Hamiltonians using $H_R(-h;\Omega_2,0,\delta)$. It can be verified that $(x,y,-z)\in EA_+(-h;\tau)$; therefore there exist $\Omega'$ and $\delta'$ such that 
    $$U(\tau;-h;\Omega',0,\delta')$$
    has Weyl chamber coordinates $(x,y,-z)$. As single-qubit gates before and after do not change the Weyl chamber coordinate, the gate
        $$(\sigma_Z\otimes I)U(\tau;-h;\Omega',0,\delta')(\sigma_Z\otimes I)=U(-\tau;h;0,\Omega',-\delta')$$
    also has Weyl chamber coordinates $(x,y,-z)$. The gate 
    $$U(\tau;h;0,\Omega',-\delta')$$
    then has Weyl chamber coordinates $(x,y,z)$ as it is conjugate with $U(-\tau;h;0,\Omega',-\delta')$, indicating that $\Omega=\Omega'$ and $\delta=-\delta'$ suffices for the AshN-EA- scheme.
\end{proof}

\subsubsection{Transforming Weyl chamber coordinates to spectrum}

The rest of the report focuses on proof of~\Cref{lem:nd} and~\Cref{lem:ea}. Before diving into the proofs, we make a few observations that help simplify the problem to be solved.

In the case of AshN gates, the problem of determining Weyl chamber coordinates can be mapped to determining the spectrum of a locally equivalent gate, or more specifically, gates of the form $\exp\{-iHt\}\cdot YY$.
We first state the intuition for this treatment. For all the AshN gates, the Hamiltonians involved come from the linear subspace spanned by $\{XX, YY, ZZ,XI, IX, ZI + IZ\}.$ It is easy to check that every entry of every element in this linear subspace is real and thus symmetric, meaning that all the AshN gates obtained by exponentiating the AshN Hamiltonians must be symmetric as well. For such a symmetric unitary $U$, its KAK decomposition looks like $(A_1\otimes A_2) \exp\{i\vec{\eta}\cdot \vec{\Sigma}\}(A_1\otimes A_2)^T$ \footnote{A rigorous proof involves transformations back and forth into the magic basis and showing that a symmetric unitary can be diagonalized by a real orthogonal matrix.}, where $A_1,A_2\in SU(2)$. For a single qubit gate $A\in SU(2)$, one can check that $Y A^T Y = A^\dag$. Therefore 
\begin{align*}
    U\cdot YY&=(A_1\otimes A_2) (\exp\{i\vec{\eta}\cdot \vec{\Sigma}\}\cdot \sigma_Y\otimes \sigma_Y)(A_1\otimes A_2)^\dag\\
    &\sim \exp\{i\vec{\eta}\cdot \vec{\Sigma}\}\cdot YY,
\end{align*} 
the spectrum of which uniquely determines the Weyl chamber coordinates $\vec{\eta}$.




\subsubsection{Overview of Proofs}
\begin{figure}[ht]
    \centering
\begin{tikzpicture}
\node[block] (a) {Problem Statement,\\ Canonicalization (\Cref{sec:ps})};  
\node[block,below left=of a] (b) {Block \\ Diagonalization (\Cref{sec:nd_canon})};   
\node[block,below right=of a] (c) {Reparametrization, \\ Block \\ Diagonalization (\Cref{sec:ea_canon})};  
\node[block,below=of b] (d) {AshN-ND scheme (\Cref{sec:nd_proof})}; 
  \node[block, below=of c] (e) {$(\alpha,\beta)\Leftrightarrow T\Leftrightarrow (x, y, z)$ (\Cref{sec:maps})};   
  \node[block, below=of e] (f) {AshN-EA scheme (\Cref{sec:ea_proof})};  
  \node[block, below left=of f] (g) {AshN scheme with Bounded Amplitudes (\Cref{sec:bounded})};
  \draw[line] (a)-- (b) node[midway, left=0.5cm] {AshN-ND};  
  \draw[line] (a)-- (c) node[midway, right=0.5cm] {AshN-EA};  
  \draw[line] (b)-- (d);  
  \draw[line] (c)-- (e); 
  \draw[line] (e)-- (f); 
  \draw[line] (f)--(g);
  \draw[line] (d) -- (g);

\end{tikzpicture}  
    \caption{Overview of the proof.}
    \label{fig:overview}
\end{figure}
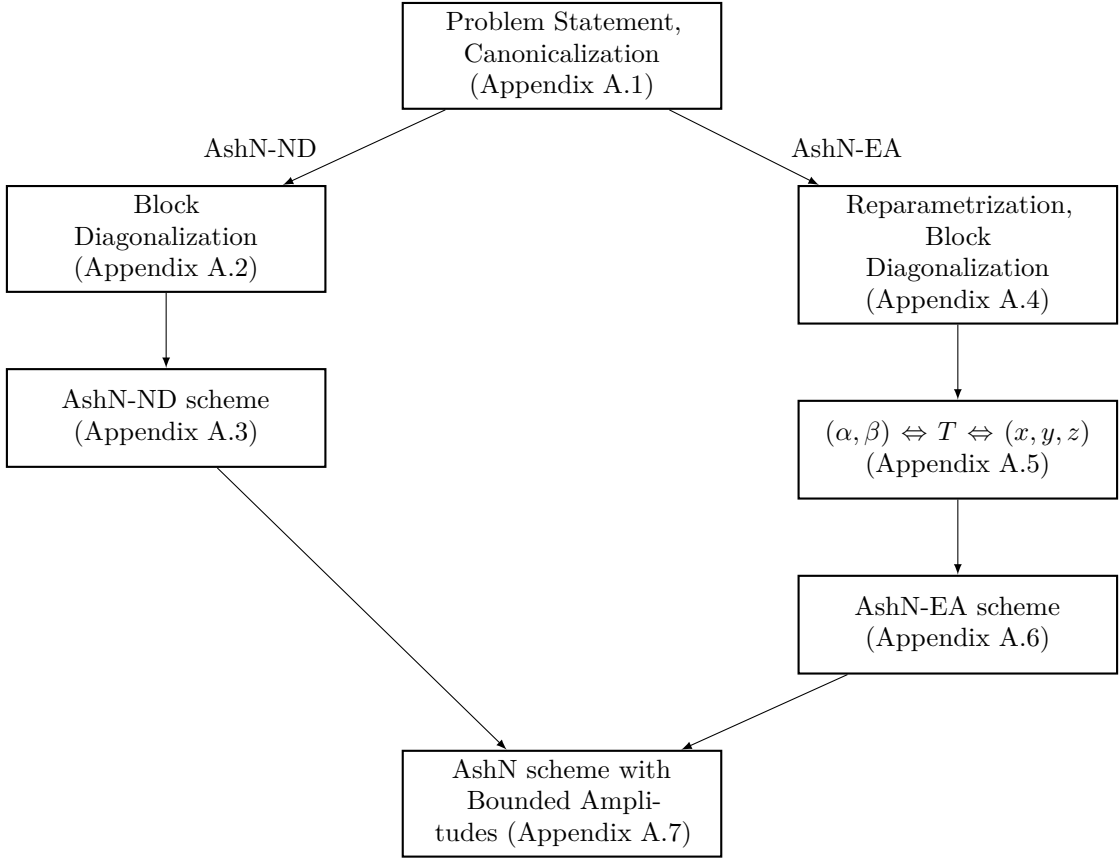
In~\Cref{fig:overview}, we draw a block diagram describing the proof of the theorems.

For the AshN-ND and AshN-ND-EXT schemes, we can show that $H_R(h;\Omega_1, \Omega_2, 0)$ can be block diagonalized and explicitly exponentiated to obtain $V(\tau;h;\Omega_1,\Omega_2,0)$. From the spectrum we obtain the Weyl chamber coordinates as a function of $\Omega_1, \Omega_2$ and show that by for $\vert \Omega_1\vert, \vert \Omega_2 \vert \leq \frac{\pi}{2\tau}$, the coordinates span $ND(h;\tau)$. This proves~\Cref{lem:nd}.

For the AshN-EA schemes, instead of directly computing the exponential, we consider the characteristic polynomial of $H_R(\tau;h;\Omega, 0, \delta)$. Other than the trivial eigenvalue -1, the remaining eigenvalues can be parameterized by $-\alpha - \beta, \alpha, 1+\beta$ for $\alpha \in [0,1]$ and $\beta \in [0,\infty)$, where we make use of the fact that $H_R$ is traceless. Using this parameterization, we can define a function $T$ that involves trace of $V(\tau;h; \Omega(\alpha,\beta), \delta(\alpha,\beta)$:
\begin{align*}
    T(\tau; h; \alpha, \beta):= \tr[V(\tau;h; \Omega(\alpha,\beta), \delta(\alpha,\beta)] + e^{i\tau}.
\end{align*}
We can show that this function $T$ along with $e^{i\tau}$ is sufficient to encode the Weyl chamber coordinates $(x,y,z)$. Hence, it is sufficient to prove that the following maps are surjective:
$$(\Omega,\delta)\in\mathbb{R}_{\geq 0}^2 \xtwoheadrightarrow{\phi_1} (\alpha,\beta)\in [0,1]\times [1,\infty) \xtwoheadrightarrow{\phi_2} T\in S \xtwoheadrightarrow{\phi_3} (x,y,z)\in EA_+(h;\tau).$$
We prove the surjectivity of $\phi_2$ and $\phi_3$ using the following respective approaches:

\begin{itemize}
    \item For $\phi_2$, we prove surjectivity using continuity: Since the map $\phi_2$ is continuous by the definition of $T$, we take a contour $C\subseteq [0,1]\otimes [1,\infty)$ in the domain and inspect its image $\phi_2(C)$. As the map is continuous, the image of the region enclosed by $C$ must contain the region enclosed by $\phi_2(C)$, which we show contains $S$.
    \item For $\phi_3$, we look at the fibers $\phi_3^{-1}(x,y,z)$. Each $(x,y,z)$ gives rise to a finite number of choices of the spectrum of $V$ due to different possible canonicalizations, which in turn gives rise to a finite number of $T$. We fix one particular canonicalization 
    $$(x,y,z)\mapsto (-e^{i(x+y+z)}, e^{i(x-y-z)}, -e^{i(-x+y-z)}, e^{i(-x-y+z)}),$$
    such that $\phi_3$ can be made invertible. Surjectivity is proven by verifying that $\phi_3^{-1}(EA_+(h;\tau))\subseteq S$.
\end{itemize}
We can then bound $\vert\Omega\vert, \vert \delta\vert$ by looking at the range of values $\alpha,\beta$ need to take on in the contour $C$. This proves~\Cref{lem:ea}. 

The uniform upper bound on the drive amplitudes can be tracked throughout the proofs. For the zero ZZ coupling case, we can analyze the gate time upper bound, average gate time, and drive strength bounds by appropriately combining $ND(0;\tau)$ with $\tau$ approaching $\pi$ to cover Weyl chamber coordinates that would normally yield a large driving amplitude. 

\subsection{Canonicalization for AshN-ND}
\label{sec:nd_canon}
The Hamiltonian corresponding to the AshN-ND scheme can be explicitly exponentiated. To see this more clearly, we conjugate the X- and Z-bases:
\begin{align*}
    H_{ND}=&(H\otimes H) H_R(h;\Omega_1,\Omega_2,0)(H\otimes H)\\=&\begin{pmatrix}\frac12 + 2\Omega_1 & && \frac{h-1}2\\&-\frac12+2\Omega_2 & \frac{1+h}2&\\&\frac{1+h}2&-\frac12-2\Omega_2&\\\frac{h-1}2&&&\frac12-2\Omega_1 \end{pmatrix}.
\end{align*}

This is a $2+2$ block-diagonal matrix which can be explicitly diagonalized. We then have
\begin{align*}
    &V(\tau;h;\Omega_1,\Omega_2,0)\\
    \sim&(H\otimes H)V(\tau;h;\Omega_1,\Omega_2,0)(H\otimes H)\\
    =&\exp\{-i H_{ND} T\}\cdot YY\\
    =&\begin{pmatrix}
    \frac{ie^{-i \tau/2}(h-1)\sin S_1\tau}{2S_1}&&&\frac{e^{-i\tau/2}(-S_1\cos S_1\tau+i\Omega_1\sin S_1\tau)}{S_1}\\&\frac{-ie^{i\tau/2}(1+h)\sin S_2\tau}{2S_2}&\frac{e^{i\tau/2}(S_2\cos S_2\tau-i\Omega_2\sin S_2\tau)}{S_2}&\\&\frac{e^{i\tau/2}(S_2\cos(S_2\tau)+i\Omega_2\sin S_2\tau)}{S_2} & \frac{-ie^{i\tau/2}(1+h)\sin S_2\tau}{2S_2} &\\ \frac{e^{-i\tau/2}(-S_1\cos S_1\tau-i\Omega_1\sin S_1\tau)}{S_1}&&&\frac{ie^{-i \tau/2}(h-1)\sin S_1\tau}{2S_1}
    \end{pmatrix}
\end{align*}
where $S_1:=\sqrt{4\Omega_1^2+(1-h)^2/4}, S_2:=\sqrt{4\Omega_2^2+(1+h)^2/4}$. The spectrum of $V$ is then $\{ie^{-i(t/2\pm \theta_1)}, -ie^{i(t/2\pm \theta_2)},\}$, where $\cos\theta_1 := \frac{(1-h)\sin(S_1 \tau)}{2S_1}$ and $\cos\theta_2 := \frac{(1+h)\sin (S_2 \tau)}{2S_2}$. It can be verified that such a gate corresponds to Weyl chamber coordinates $(\tau/2, y, z)$, where
\begin{align}
\label{eq:ypmz}
\sin(y+z)=\frac{(1+h)\sin (S_2 \tau)}{2S_2}, \sin(y-z)=\frac{(1-h)\sin (S_1 \tau)}{2S_1}.
\end{align}

It suffices to prove that the polygon $ND(h;\tau)$ can be spanned by choosing appropriate $S_1\geq \frac{1-h}2$ and $S_2\geq \frac{1+h}2$.

\subsection{Proof of~\texorpdfstring{\Cref{lem:nd}}{Lemma \ref{lem:nd}}}
\label{sec:nd_proof}
\begin{proof}
Given the definition of $ND(h;\tau)$, it suffices to prove that by varying $S_1$ and $S_2$ over appropriate ranges, $(1-h)\sin(S_1\tau)/2S_1$ and $(1+h)\sin(S_2\tau)/2S_2 $ cover appropriate ranges such that $y\pm z$ take all possible values in $ND(h;\tau)$. We will make use of the following lemma:
\begin{lem}
For $t\in [0, \pi]$, the range of $t\sin(w)/w$ for $w\in [t,\pi]$ covers $[0, \sin(t)]$.
\end{lem}
\begin{proof}
    $t\sin(w)/w$ is $\sin(t)$ when $w=t$ and is 0 when $w=\pi$. Since $t\sin(w)/w$ is continuous in $w$ for $w>0$, the range of $t\sin(w)/w $ for $w\in [t, \pi]$ covers $[0, \sin(t)]$.
\end{proof}
\noindent By re-writing the RHS of~\Cref{eq:ypmz} as $t \sin(w)/w$ for $t=\frac{1+h}2\tau$, $w = S_2\tau$ and $t=\frac{1-h}2\tau$, $w=S_1\tau$, respectively, we can let
$$S_1 \in [(1-h)/2, \pi/\tau], S_2 \in [(1+h)/2, \pi/\tau]$$ 
and the lemma implies the RHS of~\Cref{eq:ypmz} will cover $[0, \sin \frac{(1\pm h)\tau}{2}]$. According to the definition of $ND(h;\tau)$, this is exactly the range needed to cover all possible $y\pm z$. 
$S_1,S_2\leq \pi/\tau$ yields the desired bounds for $\vert \Omega_1 \vert,\vert \Omega_2\vert$ respectively.
\end{proof}

\subsection{Canonicalization \& Reparameterization for AshN-EA}
\label{sec:ea_canon}

For the AshN-EA scheme, we mainly deal with no $ZZ$ talk, i.e.\ the case $h=0$. We prove the following 
\begin{lem}[AshN-EA]
    For any $\tau \in(0, 2\pi]$, for any Weyl chamber coordinates $(x,y,z)\in EA_+(0;\tau)$, there exist $\Omega_1,\delta$ such that the interaction coefficients for $U(\tau;0;\Omega_1,0,\delta)$ are $(x,y,z)$. Moreover, $\Omega_1,\delta$ can be taken such that $\vert\Omega_1\vert,\vert\delta\vert\leq \frac \pi \tau + \frac{1}{2}$.
    \label{lem:ea_zeroh}
\end{lem}.

\Cref{lem:ea_zeroh} differs from \Cref{lem:ea} in two aspects: \Cref{lem:ea_zeroh} requires $h=0$ whereas \Cref{lem:ea} does not, and \Cref{lem:ea_zeroh} only requires $\tau\leq 2\pi$ whereas \Cref{lem:ea} requires $\tau\leq \pi$. We prove \Cref{lem:ea} from \Cref{lem:ea_zeroh} in ~\Cref{sec:nzh}. The Hamiltonian then reads
$$H_{EA}:=\frac12(XX+YY) + \Omega(XI + IX) + \delta(ZI+IZ).$$

We start by observing that all terms have a common eigenvector $(0,1,-1,0)$, and it is consequently an eigenvector of $H_{EA}$ with eigenvalue $-1$. The other three eigenvalues are difficult to solve for explicitly, making it difficult to directly exponentiate the Hamiltonian.\footnote{Technically such an explicit exponentiation can be done as the remaining characteristic polynomial is only cubic. We proceed with another approach that will prove to be simpler.} We instead look at the remaining characteristic polynomial:

\begin{align*}
    P(\lambda)&:=\frac{\det(H_{EA}-\lambda I)}{\lambda + 1}.
\end{align*}

$P$ is a cubic polynomial with leading coefficient $+1$; moreover it can be verified that $P(1)=-4\Omega^2\leq 0$, and $P(0)=4\delta^2\geq 0$. This indicates that the three real roots of $P$ lies in $(-\infty,0],[0,1]$ and $[1,+\infty)$ respectively. Moreover, the three roots should sum up to $1$ as $H_{EA}$ is traceless. We can uniquely reparameterize the eigenvalues by $-\alpha-\beta,\alpha, 1+\beta$, where $\alpha,\beta\in Q :=[0,1]\times[0,\infty)$. Inverting the parameterization by looking at the coefficients of $P(\lambda)$ gives $\Omega= \sqrt{(1-\alpha)\beta(1+\alpha+\beta)}/2$ and $\delta= \sqrt{\alpha(\alpha+\beta)(1+\beta)}/2$. Since $\Omega, \delta$ can take any non-negative value, this parameterization constitutes a bijection.

\subsection{Correspondence between \texorpdfstring{$(\alpha,\beta)$}{(\textalpha,\textbeta)}, \texorpdfstring{$T$}{T} and \texorpdfstring{$(x,y,z)$}{(x,y,z)}}
\label{sec:maps}
With the above reparameterization, we know explicitly the eigenspectrum of $H_{EA}$. We can therefore write down the following function of the trace of $V(T;0;\sqrt{(1-\alpha)\beta(1+\alpha+\beta)}/2,0, \sqrt{\alpha(1+\beta)(\beta+\alpha)}/2)$: 
\begin{align}
    T(\tau;0;\alpha,\beta)&:=\tr[V]+e^{i\tau}.
    \label{eqn:deft}
\end{align}
We claim that showing the range of $T$ covers a certain set $S \subseteq \mathbb{C}^2$ is sufficient to prove~\Cref{lem:ea}. To see this, we investigate the correspondence between the eigenvalue parameters $(\alpha,\beta)$, trace quantity $T$, and the Weyl chamber coordinates $(x,y,z)$.
Let us take a step back to revisit what we want to prove: We want to show that the Weyl chamber coordinates of $V(\tau;h;\Omega,0,\delta)$ maps the pair $(\Omega,\delta)$ surjectively onto $EA_+(h;\tau)$. To prove this, we want to find a series of three surjective maps that compose this particular map:
$$(\Omega,\delta)\in\mathbb{R}^2_{\geq 0} \xtwoheadrightarrow{\phi_1} (\alpha,\beta)\in [0,1]\times [1,\infty) \xtwoheadrightarrow{\phi_2} T\in S \xtwoheadrightarrow{\phi_3} (x,y,z)\in EA_+(h;\tau).$$
We have already worked out $\phi_1$, showing that it is a bijection. The rest of the proof focuses on finding $\phi_2,\phi_3$ as well as the set $S\subseteq\mathbb{C}^2$ such that surjectivity holds.

The map $\phi_2$ is directly defined by~\Cref{eqn:deft}, but the map $\phi_3$ is less obvious. To see why $T$ determines the Weyl chamber coordinate $(x,y,z)$ of $V$, it suffices to show that the spectrum of $V$ can be recovered from $T$. Since $V\in SU(4)$, the four eigenvalues have the form $-\exp\{i\tau\}, x_1,x_2,x_3$, where
$$\begin{cases}x_i^{-1}=\bar{x}_i, i=1,2,3,\\x_1x_2x_3 = -\exp\{-i\tau\},\\ x_1+x_2+x_3 = T.\end{cases}$$
This indicates that $x_1,x_2,x_3$ are the three roots of the cubic polynomial
\begin{align*}
    &(x-x_1)(x-x_2)(x-x_3)\\
    =&x^3 - (x_1+x_2+x_3)x^2 + x_1x_2x_3(x_1^{-1} + x_2^{-1}+x_3^{-1})x - x_1x_2x_3\\
    =&x^3 - T x^2 -\exp\{i\tau\}(\overline{(x_1+x_2+x_3)}x-1)\\
    =&x^3 - T x^2 -\exp\{i\tau\}(\bar{T}x-1)
\end{align*}
thus establishing $\phi_3$.

Before proceeding, we define $S$ to be the circular segment enclosed by the arc around the unit circle from $A:=1$ to $B:=e^{-i\tau}$ by decreasing the complex argument by $\tau$, and the line segment $\overline{AB}$. 

\subsection{Proof of \texorpdfstring{\Cref{lem:ea}}{Lemma \ref{lem:ea}}}
\label{sec:ea_proof}

\subsubsection{Surjectivity of \texorpdfstring{$\phi_2$}{\textphi 2}}
As mentioned before, we investigate the image of a contour $C\in[0,1]\times[0,\infty)$ to establish the surjectivity. We take the contour to consist of four line segments:
$$C=(0,0)\rightarrow (0,2\pi/\tau - 1)\rightarrow (1,2\pi/\tau)\rightarrow (1,0)\rightarrow (0,0).$$

When $\alpha,\beta$ traverses along $C$, we claim $T$ encloses $S$. We compute
\begin{align*}
    T|_{\alpha=0}&=1 + \frac{\sin((1+\beta)\tau/2)}{(1+2\beta )\sin(\tau/2)}(e^{-i\tau}-1)\supseteq \overline{AB},\\
    T|_{\alpha=1}&=e^{-i\tau} = B,\\
    T|_{\beta=0}&=e^{-i\tau} = B.
\end{align*}
A special case is when $\tau=2\pi$, where $A=B$ and $T|_{\alpha=0}=A=B$.

\begin{figure}
\centering
\begin{tikzpicture}[scale=2.3]

\def\ang{1}
\def\R{1}
\coordinate (B) at (-120:1);
\coordinate (A) at (1,0);
\coordinate (O) at (0,0);
\node [below right] at (O) {$0$};
\node [below right] at (A) {$A$};
\node [below left] at (B) {$B$};
\draw (1.3, 0.17) node [below right] {$P(0)$};

\filldraw[fill=green!20!white, draw=green!50!black]
    (1,0) -- (-120:1) arc (-120:0:1)-- cycle;
\node  at (-60:0.75) {$S$}; 
\draw [blue, thick,  domain=0:2, samples=400] 
 plot ({1-1.5*(sin(120 * (\x + 0.5))/((1+2*\x) * sin(60)))}, {-sqrt(3)/2*(sin(120 * (\x + 0.5))/((1+2*\x) * sin(60)))} );
\draw [red, thick,  domain=0:1, samples=400] 
 plot ({(cos(4*180*\x/3)-sqrt(3)*sin(4*180*\x/3) + (1+\x) * (4 * (3 + 2*\x) * cos(2 * 180 * \x / 3) + \x * sin(1/6*180*(-1 + 8 * \x))))/((2+3*\x)* (5+3*\x))}, {(sqrt(3)*cos(4*180*\x/3)+sin(4*180*\x/3)+ (1+\x) * (-4 * (3 + 2*\x) * sin(2 * 180 * \x / 3) - \x * sin(1/3*180*(1 + 4 * \x))))/((2+3*\x)* (5+3*\x))} );
\draw[
        step=.5cm,
        gray,
        very thin
        ] 
        (-1.4,-1.4) grid (1.4,1.4);
\draw[->] (-1.5,0) -- (1.5,0) coordinate (x axis);
\draw[->] (0,-1.5) -- (0,1.5) coordinate (y axis);
\draw[dashed] (A) -- ++(-150:2.5);
\draw[dashed] (A) -- ++(30:0.5);
\node [right]at (1.5,0)(x){$1$};    
\node [above]at (0,1.5){$i$};
\draw (0,0) circle [
                    radius=1cm
                    ];

\end{tikzpicture}
\caption{Illustration of $\phi_2(C)$ and $S$ for $\tau = 2\pi/3$. The blue line corresponds to $\alpha=0$, and the red curve corresponds to the trajectory of $P(\alpha)$. Notice that $P(0)$ is on the side of $A$, and $P(\alpha)$ lies on the same side of the line $\overline{AB}$, and the modulus is always lower bounded by $1$. Those three properties ensure that it encloses $S$ with the blue line.}
\label{fig:illu}
\end{figure}
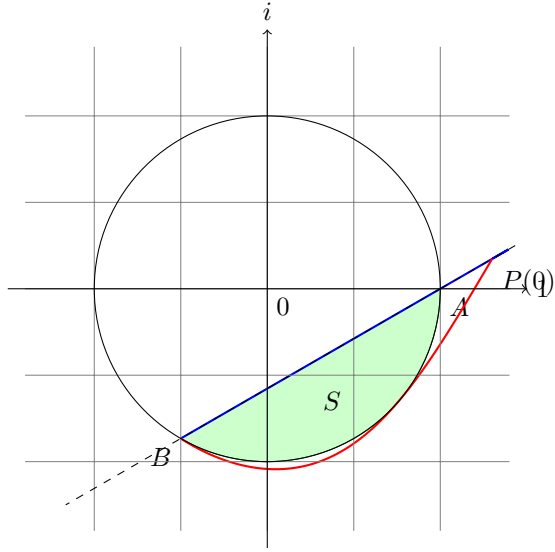

We now take a closer look at the trajectory $P(\alpha)=T|_{\beta = 2\pi/\tau - 1 + \alpha}$. To show $\phi_2(C)$ encloses $S$, we show the following:
\begin{itemize}
    \item $A$ lies between $B$ and $P(0)$. This is because
    $$\frac{\sin((1+\beta)\tau/2)}{(1+2\beta )\sin(\tau/2)}=-\frac{\sin\alpha\tau/2}{(1+2\beta)\sin(\tau/2)}\leq 0.$$
    \item $P(\alpha)$ always lie on the same side of $\overline{AB}$ as $\arctikz{AB}$, or equivalently,     \begin{align}
    R(\alpha):=&\Re[(P(\alpha) - 1) e ^{i\tau / 2}]\\
    =&2\sin[\frac{\tau\alpha}{2}]\left(\sin(\frac{\tau(1-\alpha)}2)-\frac{\tau(2\pi-t(1-\alpha))(1-\alpha)}{(2\pi+\tau(-1+3\alpha))(4\pi+\tau(-1+3\alpha))}\sin(\frac{\tau(1-3\alpha)}2)\right)\geq 0.
\end{align}

Let $A:=\frac{\tau(2\pi-\tau(1-\alpha))(1-\alpha)}{(2\pi+\tau(-1+3\alpha))(4\pi+\tau(-1+3\alpha))}$. It can be proven that $A\in[0,1]$ when $\tau\leq 2\pi$ and $A\leq \frac{\tau(1-\alpha)}{4\pi}$ when $\alpha\geq 2/3$. Then
\begin{itemize}
    \item When $\tau(1-3\alpha)/2\geq 0$, we have $0 \leq \sinc(\pi-\frac{\tau(1-3\alpha)}{2})\leq \sinc(\pi-\frac{\tau(1-\alpha)}{2})$ since $\sinc(x)$ decreases monotonically in $[0,\pi]$. Together with $0\leq \tau(1-\alpha)\leq 2\pi \leq (4\pi+\tau(-1+3\alpha))$ can we get $R(\alpha)\geq 0$;
    \item When $\tau(1-3\alpha)/2\in[-\pi,0]$, $\sin(\frac{\tau(1-3\alpha)}{2})\leq 0$ and $A\geq 0$. We have $R(\alpha)\geq 0$;
    \item When $\tau(1-3\alpha)/2\leq -\pi$, we have $\alpha\geq 2/3$ given $\tau\leq 2\pi$. Then $\tau(1-\alpha)/2\leq \pi/3$ and $\sin(\tau(1-\alpha)/2)\geq \frac{\tau(1-\alpha)}{4\pi}$, whereas $A\sin(\frac{\tau(1-3\alpha)}{2})\leq \frac{\tau(1-\alpha)}{4\pi}$. We again have $R(\alpha)\geq 0$.
\end{itemize}
\item $|P(\alpha)|\geq 1$. Indeed, we have
$$|P(\alpha)|^2 - 1=\frac{16\tau(2\pi-(1-\alpha)\tau)(1-\alpha)(\pi+\tau\alpha)(2\pi+(2\alpha-1)\tau)\sin^2[\frac{\tau(1-3\alpha)}{2}]}{(2\pi+\tau(-1+3\alpha))^2(4\pi+\tau(-1+3\alpha))^2}\geq 0$$
for $\tau \leq 2\pi$.
\end{itemize}
\Cref{fig:illu} illustrates how the three conditions suffice for proving $\phi_2(C)$ encloses $S$.

\subsubsection{Surjectivity of \texorpdfstring{$\phi_3$}{\textphi 3}}
\label{subsec:phi3_surj}
Assume $\tau \in [0,2\pi]$. We claim that the image of $W\supseteq EA_+(h;\tau)$ defined as the triangle $PQR$, where
\begin{align*}
    P &:=(\tau/2,\tau/2,0),\\
    Q &:= (\tau/2,\tau/4,\tau/4) ,\\
    R &:= (\tau/3,\tau/3,\tau/3)
\end{align*}
via $\phi_3^{-1}$ (with respect to our fixed canonicalization) is exactly $S$\footnote{Note that $W$ is not necessarily contained in the Weyl chamber whereas $\phi_3^{-1}$ can nevertheless be extended to $\mathbb{R}^3$.}. Now, $\phi_3^{-1}$ maps a coordinate $(x,y,z)$ to the complex number
\begin{align}
\label{eq:weyl_trace}
    e^{i(x-y-z)} - e^{i(-x+y-z)} + e^{i(-x-y+z)} .
\end{align}
Thus,
\begin{align*}
   \phi_3^{-1}: P & \mapsto e^{-i\tau} = B\\
   Q &\mapsto 1= A\\
   R &\mapsto e^{-i\tau/3} 
\end{align*}
We consider the image of $\overline{PQ}$ parameterized by $(\tau/2, (2-t)\tau/4, t\tau/4)$, $t\in[0,1]$:
\begin{align*}
    \phi_3^{-1}(\tau/2, (2-t)\tau/4, t\tau/4)=1-2i\sin \frac{(1-t)\tau}{2} e^{-i\tau/2}.
\end{align*}
This is clearly $\overline{AB}$. 
We next look at the image of $\overline{QR}$, parameterized by $(\tau/2-t\tau/6,\tau/4+t\tau/12, \tau/4+t\tau/12)$, $t\in[0,1]$:
\begin{align*}
    \phi_3^{-1}(\tau/2-t\tau/6,\tau/4+t\tau/12, \tau/4+t\tau/12)=\exp\{-it\tau/3\}.
\end{align*}
We also look at the image of $\overline{RP}$, parameterized by $(\tau/3+t\tau/6,\tau/3+t\tau/6,(1-t)\tau/3)$, $t\in[0,1]$:
\begin{align*}
    \phi_3^{-1}(\tau/3+t\tau/6,\tau/3+t\tau/6,(1-t)\tau/3)=\exp\{-i(1+2t)\tau/3\}.
\end{align*}
Hence, we can see explicitly that this traces out $\arctikz{AB}$. Therefore $\partial W \mapsto \partial S$. Now, $W$ is a compact set in $\R^3$. Since~\Cref{eq:weyl_trace} is continuous, the image of $W$ is a compact set in $\C$. Since $T \leq 2\pi$, $\partial S$ is the boundary of a triangle with one of the edges replaced by an arc. 
Hence, $\partial S$ is a simple closed curve, and so by the Jordan curve theorem, $S$ is the only compact set with boundary $\partial S$. The claim follows. Bounds on $\Omega_1$ and $\delta$ can be derived from $\beta\leq 2\pi/\tau-1+
\alpha$ and $\alpha\in[0,1]$.

\subsubsection{AshN-EA for nonzero \texorpdfstring{$h$}{h}}
\label{sec:nzh}
We finally deal with the case $h\neq 0$. Note that $[XX+YY+ZZ, H_R(h;\Omega_1,0,\delta)]=0$; this means we can separate the Hamiltonian into two parts
$$H_R(h;\Omega_1,0,\delta) = h/2(XX+YY+ZZ) + (1-h)H_R(0, \Omega_1/(1-h), 0,\delta/(1-h))$$
and exponentiate each part separately:
$$V(\tau;h;\Omega,0,\delta)=V((1-h)\tau; 0; \Omega/(1-h),0,\delta/(1-h))\cdot\exp\{-ih\tau/2(XX+YY+ZZ)\},$$
the Weyl chamber coordinate of which (before canonicalization) covers the triangle $P'Q'R'= EA_+(h;\tau)$,\footnote{This is the triangle $PQR$ for time $(1-h)\tau\leq 2\pi$ shifted by $(h\tau/2,h\tau/2,h\tau/2)$, and the proof follows that of~\Cref{subsec:phi3_surj}.} where
\begin{align*}
    P'&:=(\tau/2,\tau/2, h\tau/2),\\
    Q'&:=(\tau/2, (1+h)\tau/4, (1+h)\tau/4),\\
    R'&:=((2+h)\tau/6, (2+h)\tau/6, (2+h)\tau/6).
\end{align*}

\subsection{AshN Scheme with Bounded Amplitudes}
\label{sec:bounded}
So far we have established that the AshN-ND scheme and the AshN-EA schemes give rise to optimal-time realization of any Weyl chamber coordinates. However, the amplitudes and detuning required are on the order of $\Omega(1/\tau)$. Thus, for infinitesimal gate times, the amplitudes and detuning required can be unbounded. We here incorporate the AshN-ND-EXT scheme to relax this unphysical requirement without substantially increasing the gate time.

First note that the interaction time $\tau$ in ~\Cref{lem:nd,lem:ea,cor:ea} are all upper bounded by $\pi$. If we combine the AshN-ND-EXT strategy with the AshN-ND, AshN-EA+ and AshN-EA- as stated in the main body, all the gates can be realized within time $\pi$ and strength $(1+|h|)\pi/\tau + 1/2$. It can be verified that for all $r\leq (1-|h|)\pi/2$, every Weyl chamber can be realized with one of the four subschemes. If we take $r\leq(1-|h|)\pi/2$ we get a uniform upper bound of strengths of $\frac{2(1+|h|)}{1-|h|}+\frac12$.

\subsubsection{Average gate time as a function of the cutoff parameter \texorpdfstring{$r$}{r}}

When $h=0$, we described in the main body a hybrid compilation scheme of AshN-ND-EXT and AshN-EA or AshN-ND involving a hyperparameter $r\in[0,\frac\pi2]$ we call the \emph{cutoff}. We recall the protocol to implement the interaction coefficients $(x,y,z)$:
\begin{itemize}
    \item If $\tau_{opt}\leq r$, we adopt the AshN-ND-EXT scheme to realize the gate in time $\pi-2x$;
    \item Otherwise, we adopt AshN-ND, AshN-EA+ or AshN-EA- depending on which sector $(x,y,z)$ lies in and realize the gate within the optimal time.
\end{itemize}
It is easy to see that such a scheme realizes all gates with 
$$\max\{|\Omega_1+\Omega_2|,|\Omega_1-\Omega_2|,|\delta|\}\leq \frac{\pi}r + \frac{1}{2}.$$
For $r=0$ we recover the optimal time AshN compilation scheme. However, for any $r>0$, the maximum gate time approaches a non-optimal $\tau$ when $(x,y,z)$ approaches $(0,0,0)$. We show below that the \emph{average} gate time, on the other hand, quickly approaches the optimal one without significantly increasing the required amplitudes.

Formally we define the required gate time $\mathcal{T}$ given the Weyl chamber coordinate $(x,y,z)$ and the cutoff parameter $r$ as
$$\mathcal{T}(x,y,z;r):=\begin{cases}
    \max\{2x, x+y+|z|\}, \text{if $\max\{2x, x+y+|z|\}\geq r$},\\
    \pi-2x,\text{ otherwise,}
\end{cases}$$
and the average gate time with cutoff $r$
$$\mathcal{T}_{avg}(r):=\int \mathcal{T}(x,y,z;r) d\mu_W,$$
where $\mu_W$ is the induced measure on the Weyl chamber induced by the Haar measure on $SU(4)$ defined as~\cite{watts2013metric}
$$d\mu_W:=\frac{48}\pi \sin(x+y)\sin(x-y)\sin(y+z)\sin(y-z)\sin(x+z)\sin(y-z)dxdydz.$$
We can explicitly compute this: 
\begin{align*}
    \mathcal{T}_{avg}(r)=& \frac1{28800\pi}(225(-176 r^2+96\pi r-105)\cos (4 r)+50(-576 r^2+576\pi r-30\cos (6 r)+252\pi^2+97)\\
    +&60 (480 (\pi-2 r)\sin (r)-603 (\pi-2 r)\sin (2 r)-128 (\pi-2 r)\sin (3 r)+30 (19\pi-33 r)\sin (4 r)\\
    -&480 (\pi-2 r)\sin (5 r)+65 (\pi-2 r)\sin (6 r))-59049\cos(\frac{4 r}{3})+51708\cos (2 r)+9216\cos (3 r)\\
    +&15360\cos (5 r))\\
    =&\frac{-76+315\pi^2}{720\pi} + \frac{2213}{5040}r^9 - \frac{160303}{204120\pi}r^{10} + O(r^{11}).
\end{align*}

\section{Proof of Circuit Synthesis Results}\label{app:decomp}
The following is an outline of a mathematical proof that a generic $SU(8)$ element can be decomposed into 11 two-qubit gates. 

\subsection{Revisit two-qubit gate decomposition from the Hamiltonian perspective}
We first revisit the KAK decomposition of a generic two-qubit gate but from a Hamiltonian perspective. Take $U\in SU(4)$ and define $H:=i\log (U)$ as the corresponding Hamiltonian. By the definition of $U$, $H$ is a traceless Hermitian matrix. We can express it in the Pauli basis to get the following matrix:
$$O(H)=\begin{bmatrix}0&H_{IX}&H_{IY}&H_{IZ}\\H_{XI}&H_{XX}&H_{XY}&H_{XZ}\\H_{YI}&H_{YX}&H_{YY}&H_{YZ}\\H_{ZI}&H_{ZX}&H_{ZY}&H_{ZZ}\end{bmatrix}\in\mathbb{R}^{4\times 4}.$$

We want to append single-qubit gates before and after the two-qubit gate to try to reduce $H$ to a certain canonical form. We here outline the procedure; the rigorous proof is given in~\cite{bennett2002optimal}. We can multiply $U$ with the single qubit terms $\exp\{-i(H_{IX} \cdot I\otimes X + H_{IY}\cdot I\otimes Y+H_{IZ}\cdot I\otimes Z + H_{XI}\cdot X\otimes I + H_{YI}\cdot Y\otimes I + H_{ZI}\cdot Z\otimes I)\}$. Although doing so may not eliminate the single-qubit Pauli terms of $H$ entirely, it should at least shrink the single-qubit terms in magnitude. By iterating this procedure, which is allowed since the single-qubit gates can be merged with previously appended ones, we should be able to eliminate entirely the single-qubit terms and get 
$$O(H')=\begin{bmatrix}0&0&0&0\\0&H'_{XX}&H'_{XY}&H'_{XZ}\\0&H'_{YX}&H'_{YY}&H'_{YZ}\\0&H'_{ZX}&H'_{ZY}&H'_{ZZ}\end{bmatrix}\in\mathbb{R}^{4\times 4}.$$

We can then apply conjugations to each qubit. This translates directly to conjugations of the corresponding Hamiltonian. The action of $SU(2)$ conjugation on the first and the second qubit corresponds respectively to left and right $SO(3)$ multiplications of the nonzero block of $O(H')$. By the local isomorphism between $SU(2)$ and $SO(3)$, such conjugations allow us to perform SVD on the nonzero block and get
$$O(H'')=\begin{bmatrix}0&0&0&0\\0&H''_{XX}&0&0\\0&0&H''_{YY}&0\\0&0&0&H''_{ZZ}\end{bmatrix}\in\mathbb{R}^{4\times 4},$$
which is exactly the KAK decomposition.

\subsection{Generalizing to three qubits}
We can apply similar arguments to finding a canonical form of $SU(8)$ modulo $SU(2)\otimes SU(4)$. We can write the Pauli coefficients of $H$ as a $4\times 16$ real matrix with $H_{I,II}=0$. Appending $SU(2)\otimes SU(4)$ on one side eliminates the first row and the first column. The problem is then reduced to the analysis of conjugations of $SU(2)\otimes SU(4)$ on the remaining $3\times 15$ matrix.

The $SU(2)$ action is fairly well understood via the local isomorphism between $SU(2)$ and $SO(3)$. However the adjoint representation $SU(4)\rightarrow SO(15)$ is not surjective from a dimension counting argument (in fact $SU(4)\cong SO(6)$). From a dimension counting perspective, conjugation by $SU(2)\otimes SU(4)$ should be able to eliminate $15+3=18$ degrees of freedom, leaving $27$ in the canonical form.

\subsection{Recursive analytical circuit synthesis with generic SU(4) gates}
We consider circuit synthesis using general single- and two-qubit gates chosen from the special unitary group $SU(4)$. In particular, we prove the following results using a recursive construction.
\begin{thm}\label{thm:3QubitRec}
Any three-qubit gate can be decomposed into a sequence of at most $11$ two-qubit generic gates chosen from $SU(4)$ and single-qubit gates. 
\end{thm}

We provide the proof for the above theorem in~\Cref{sec:proof3QubitRec}. The circuit synthesis algorithms~\cite{shende2004smaller,shende2004minimal,barenco1995elementary,cybenko2001reducing,aho2003design,knill1995approximation,mottonen2004quantum,vartiainen2004efficient} focus on decomposition using single-qubit gates and CNOT gates, which is taken to be the native gate set. The best known analytic result requires $20$ CNOT gates to compile arbitrary $3$-qubit quantum gates~\cite{shende2005synthesis}. Compared with this result,~\Cref{thm:3QubitRec} provides a decomposition with approximately half of the two-qubit gate count if we allow arbitrary two-qubit gates instead of CNOT gates. Furthermore, using this decomposition strategy for three-qubit gates, we can further derive the following theorem concerning the generic two-qubit gate count for arbitrary $n$-qubit circuit synthesis.
\begin{thm}\label{thm:CircSynRec}
An arbitrary $n$-qubit gate can be implemented in a circuit containing no more than $\frac{23}{64}4^n-\frac{3}{2}2^n$ generic two-qubit gates.
\end{thm}
We prove this in~\Cref{sec:proofCircSynRec}. Compared with the strategy using CNOT gates that gives a $\frac{23}{48}4^n-\frac{3}{2}2^n+\frac{4}{3}$~\cite{shende2005synthesis}, our decomposition provides a gate count with a one-quarter reduction.

\subsubsection{Three-qubit gate decomposition}\label{sec:proof3QubitRec}
We provide the proof for~\Cref{thm:3QubitRec}. We start by introducing some notation:
\begin{itemize}
    \item $\Qcircuit @C=1em @R=.7em {
    & {/} \qw & \gate{\phantom{U}} & \qw 
    }\qquad$ A generic gate.
    \item $\Qcircuit @C=1em @R=.7em {
    & {/} \qw & \gate{\Delta} & \qw 
    }\qquad$ A generic diagonal gate.
    \item $\Qcircuit @C=1em @R=.7em {
    & \qw & \gate{R_k} & \qw 
    }\qquad$ A single-qubit rotation gate with a generic angle along the direction $k=x,y,z$.
\end{itemize}
\noindent Another important type of gate that requires further explanation is the multiplexor. We say a gate is a quantum multiplexor with select qubits if it preserves the reduced state on select qubits. We denote the multiplexor as 
\begin{align}
\Qcircuit @C=1em @R=.7em {
& {/} \qw & \gate{}\qwx[1] & \qw & \qw\\
& {/} \qw & \gate{\phantom{U}}  & \qw & \qw
}\nonumber
\end{align}
The select qubits are on top. If there is only a single select qubit and it is most significant, the matrix of the quantum multiplexor is block diagonal:
\begin{align}
U=\begin{pmatrix}
U_0 & \\
& U_1
\end{pmatrix}.\nonumber
\end{align}
In the following, we prove the following lemma concerning the decomposition of an arbitrary multiplexor.
\begin{lem}\label{lem:3QubitMult}
Any three-qubit multiplexor with a single select qubit can be implemented using $5$ two-qubit gates, among which three are diagonal.
\end{lem}
\begin{proof}
We will prove the following decomposition:
\begin{align}
\Qcircuit @C=1em @R=.9em {
&  \qw & \gate{}\qwx[1] & \qw & \qw\\
&  \qw & \multigate{1}{\phantom{U}}  & \qw & \qw\\
&  \qw & \ghost{\phantom{U}}  & \qw & \qw
}
\qquad
\Qcircuit @C=1em @R=.9em {
& \qw & \multigate{1}{\Delta} & \qw & \multigate{1}{\Delta} &  \gate{\Delta}{2}\cwx[2] & \qw & \qw\\
=\qquad & \qw & \ghost{\Delta}  & \multigate{1}{V_1} & \ghost{\Delta} & \qw & \multigate{1}{V_2} & \qw\\
& \qw & \qw & \ghost{V_1} & \qw & \gate{\Delta} & \ghost{V_2} & \qw
}\nonumber
\end{align}
Here, $V_1$ and $V_2$ are two generic two-qubit gates and the remaining three gates are generic diagonal gates. Without loss of generality, we denote these three gates $D_1$, $D_2$, and $D_3$ from left to right. We also denote the target three-qubit gate as $U$. According to the definition of a multiplexor, we can write $U$ as
\begin{align}
\label{eq:subsystem_u}
U=\ket{0}\bra{0}\otimes U_0+\ket{1}\bra{1}\otimes U_1,
\end{align}
where $U_0,U_1\in\mathbb{C}^{4\times 4}$. We consider the three diagonal gates $D_1$, $D_2$, and $D_3$. Each of the diagonal gates contains three free parameters. However, we can decompose a generic diagonal gate into two single-qubit rotations and a diagonal gate containing one parameter (i.e., a ZZ rotation). Since $U$ is a unitary, $U_0$ and $U_1$ are unitaries. Via such a decomposition, by choosing suitable single qubit rotations, we can describe each diagonal gate by a single parameter: $D_1=ZZ(\theta_1)$, $D_2=ZZ(\theta_2)$, and $D_3=ZZ(\theta_3)$.

To prove that this circuit can compile the target unitary $U$, we only need to prove that by choosing proper parameters $\theta_i$ and generic two-qubit gates $V_1$ and $V_2$, we can construct
\begin{align}
U_0&\sim(R_z(\theta_1)\otimes I)V_1(R_z(\theta_2)\otimes R_z(\theta_3))V_2,\nonumber\\
U_1&\sim(R_z(-\theta_1)\otimes I)V_1(R_z(-\theta_2)\otimes R_z(-\theta_3))V_2.\nonumber
\end{align}
Here, $\sim$ hides a global phase and single-qubit gates. Since we can choose any $V_2$, it is sufficient to find $V_1$ and parameters $\theta_i$ such that
\begin{align}
\label{eq:u0u1dag}
U_0U_1^\dagger=(R_z(\theta_1)\otimes I)V_1(R_z(2\theta_2)\otimes R_z(2\theta_3))V_1^\dagger(R_z(\theta_1)\otimes I).
\end{align}
or
\begin{align}
V_1(R_z(2\theta_2)\otimes R_z(2\theta_1))V_1^\dagger=(R_z(-\theta_1)\otimes I)U_0U_1^\dagger (R_z(-\theta_1)\otimes I),\nonumber
\end{align}
Notice that $R_z(2\theta_2)\otimes R_z(2\theta_3)$ has four eigenvalues $e^{\pm i(\theta_2\pm\theta_3)}$, which is shared by $V_1(R_z(2\theta_2)\otimes R_z(2\theta_3))V_1^\dagger$. Hence, we only need to show that by choosing proper $\theta_1$, 
\begin{align}\label{eq:DefUp}
U':=(R_z(-\theta_1)\otimes I)U_0U_1^\dagger (R_z(-\theta_1)\otimes I)=V_1R_z(2\theta_2)\otimes R_z(2\theta_3)V_1^\dagger
\end{align}
has two pairs of conjugate eigenvalues. As \Cref{eq:u0u1dag} shows that $U_0U_1^\dagger$ has a unit determinant, the characteristic polynomial of $U'$ has the following form as it also has a unit determinant and unit eigenvalues:
\begin{align}\nonumber
\det(U'-\lambda I)=\lambda^4+\tr(U')\lambda^3+r_2\lambda^2-\tr(\overline{U'})\lambda+r_0.
\end{align}
where $r_0,r_2\in\mathbb{R}$. To prove that we can express $U'$ in the form of \Cref{eq:DefUp}, we prove the roots are two pairs of conjugate numbers. This holds if the coefficients of this characteristic polynomial are all real, which is equivalent to showing $\tr(U')$ can be made real by properly choosing $\theta_1$. Denote the diagonal elements of $U_0U_1^\dagger$ by $(U_0U_1^\dagger)_{11},(U_0U_1^\dagger)_{22},(U_0U_1^\dagger)_{33},(U_0U_1^\dagger)_{44}$ and let $(U_0U_1^\dagger)_{11}+(U_0U_1^\dagger)_{33}=r_ae^{i\theta_a}$ and $(U_0U_1^\dagger)_{22}+(U_0U_1^\dagger)_{44}=r_be^{i\theta_b}$ for some $r_a,r_b\in \mathbb{R}$ and $\theta_a,\theta_b\in[0,\pi]$. Then we have $\tr(U')=r_ae^{i(\theta_a+\theta_1)}+r_be^{i(\theta_b-\theta_1)}$. We can always pick $\theta_1$ to satisfy 
$$r_a\sin(\theta_a+\theta_1)+r_b\sin(\theta_b-\theta_1)=0$$
which is equivalent to $\tr(U')\in\mathbb{R}$. This completes the proof of~\Cref{lem:3QubitMult}.
\end{proof}

To prove~\Cref{thm:3QubitRec}, we use the following Cosine-Sine Decomposition~\cite{bullock2004note,mottonen2004quantum,tucci1999rudimentary,vartiainen2004efficient}. This decomposition enables us to split an arbitrary three-qubit unitary into a sequence of multiplexors. Formally, it can be written as
\begin{align}\label{eq:SinCosDecomp}
\Qcircuit @C=1em @R=.7em {
& \qw & \multigate{1}{\phantom{U}} & \qw\\
& {/} \qw & \ghost{\phantom{U}} & \qw
}=\Qcircuit @C=1em @R=.7em {
& \qw & \gate{} \qwx[1] & \gate{R_y} & \gate{} \qwx[1] & \qw\\
& {/} \qw & \gate{\phantom{U}} & \gate{} \qwx[-1] & \gate{\phantom{U}} & \qw
}
\end{align}
Here, the middle gate is a multiplexed $y$ rotation, which means each subsystem unitary (such as $U_0$ and $U_1$ in~\Cref{eq:subsystem_u}) is a $y$ rotation. For the case of three-qubit gates, Ref.~\cite{shende2005synthesis} provides the following decompositions for such a gate:
\begin{align}\nonumber
\Qcircuit @C=1em @R=.7em {
& \gate{R_y} & \qw\\
& \gate{} \qwx[-1] & \qw\\
& \gate{} \qwx[-1] & \qw
}=\Qcircuit @C=1em @R=.7em {
& \gate{R_y} & \targ & \gate{R_y} & \targ & \qw\\
& \gate{} \qwx[-1] & \qw & \gate{} \qwx[-1] & \qw & \qw\\
& \qw  & \ctrl{-2} & \qw & \ctrl{-2} & \qw
}=\Qcircuit @C=1em @R=.7em {
& \gate{R_y} & \ctrl{2} & \gate{R_y} & \ctrl{2} & \qw\\
& \gate{} \qwx[-1] & \qw & \gate{} \qwx[-1] & \qw & \qw\\
& \qw  & \ctrl{-2} & \qw & \ctrl{-2} & \qw
}
\end{align}
We can also reverse the order of the gates simply by taking the conjugate transpose. We then apply~\Cref{lem:3QubitMult} to the Sine-Cosine Decomposition in \Cref{eq:SinCosDecomp}:
\begin{align}
\Qcircuit @C=1em @R=.7em {
& \multigate{1}{\Delta} & \qw & \multigate{1}{\Delta} &  \sgate{\Delta}{2} & \qw & \ctrl{2} & \gate{R_y} & \ctrl{2} & \gate{R_y} & \multigate{1}{\Delta} & \qw & \multigate{1}{\Delta} &  \sgate{\Delta}{2} & \qw & \qw\\
& \ghost{\Delta}  & \multigate{1}{V_1} & \ghost{\Delta} & \qw & \multigate{1}{V_2} & \qw & \gate{} \qwx[-1] & \qw & \gate{} \qwx[-1] & \ghost{\Delta}  & \multigate{1}{V_1} & \ghost{\Delta} & \qw & \multigate{1}{V_2} & \qw\\
& \qw & \ghost{V_1} & \qw & \gate{\Delta} & \ghost{V_2} & \ctrl{-2} & \qw  & \ctrl{-2} & \qw & \qw & \ghost{V_1} & \qw & \gate{\Delta} & \ghost{V_2} & \qw
}\nonumber
\end{align}
Since the rightmost gate of the implementation of controlled-y rotation and the leftmost gate of the right multiplexor can be realized using a single two-qubit gate, we can simply the circuit as follows:
\begin{align}
\Qcircuit @C=1em @R=.7em {
& \multigate{1}{\Delta} & \qw & \multigate{1}{\Delta} &  \sgate{\Delta}{2} & \qw & \ctrl{2} & \gate{R_y} & \ctrl{2} & \multigate{1}{\phantom{U}} & \qw & \multigate{1}{\Delta} &  \sgate{\Delta}{2} & \qw & \qw\\
& \ghost{\Delta}  & \multigate{1}{V_1} & \ghost{\Delta} & \qw & \multigate{1}{V_2} & \qw & \gate{} \qwx[-1] & \qw & \ghost{\phantom{U}}  & \multigate{1}{V_1} & \ghost{\Delta} & \qw & \multigate{1}{V_2} & \qw\\
& \qw & \ghost{V_1} & \qw & \gate{\Delta} & \ghost{V_2} & \ctrl{-2} & \qw  & \ctrl{-2} & \qw & \ghost{V_1} & \qw & \gate{\Delta} & \ghost{V_2} & \qw
}\nonumber
\end{align}
Next, we observe that the first six gates are a multiplex with the first qubit selected. Thus, we can use~\Cref{lem:3QubitMult} and reverse the order to further simply the circuit into
\begin{align}
\Qcircuit @C=1em @R=.7em {
& \qw & \sgate{\Delta}{2} & \multigate{1}{\Delta} &  \qw & \multigate{1}{\Delta} & \gate{R_y} & \ctrl{2} & \multigate{1}{\phantom{U}} & \qw & \multigate{1}{\Delta} &  \sgate{\Delta}{2} & \qw & \qw\\
& \multigate{1}{V_2}  & \qw & \ghost{\Delta} & \multigate{1}{V_1} & \ghost{\Delta} & \gate{} \qwx[-1] & \qw & \ghost{\phantom{U}}  & \multigate{1}{V_1} & \ghost{\Delta} & \qw & \multigate{1}{V_2} & \qw\\
& \ghost{V_2}  & \gate{\Delta} & \qw & \ghost{V_1} & \qw & \qw & \ctrl{-2} & \qw & \ghost{V_1} & \qw & \gate{\Delta} & \ghost{V_2} & \qw
}   \nonumber
\end{align}
By combining the fifth and the sixth gate, we finally obtain a decomposition of arbitrary three-qubit gates using $11$ two-qubit gates as follows:
\begin{align}
\Qcircuit @C=1em @R=.7em {
& \qw & \sgate{\Delta}{2} & \multigate{1}{\Delta} &  \qw & \multigate{1}{\phantom{U}} & \ctrl{2} & \multigate{1}{\phantom{U}} & \qw & \multigate{1}{\Delta} &  \sgate{\Delta}{2} & \qw & \qw\\
& \multigate{1}{V_2}  & \qw & \ghost{\Delta} & \multigate{1}{V_1} & \ghost{\phantom{U}} & \qw & \ghost{\phantom{U}}  & \multigate{1}{V_1} & \ghost{\Delta} & \qw & \multigate{1}{V_2} & \qw\\
& \ghost{V_2}  & \gate{\Delta} & \qw & \ghost{V_1} & \qw & \ctrl{-2} & \qw & \ghost{V_1} & \qw & \gate{\Delta} & \ghost{V_2} & \qw
}   \nonumber
\end{align}
thereby completing the proof of~\Cref{thm:3QubitRec}.

\subsubsection{The gate count for circuit synthesis}\label{sec:proofCircSynRec}
Now, we consider the number of two-qubit gates required for generic $n$-qubit circuit synthesis. Starting from the CSD decomposition, we can obtain the following quantum Shannon decomposition~\cite{shende2005synthesis} which splits the $n$-qubit circuit into $n-1$-qubit circuits and multiplexor:
\begin{align}
\Qcircuit @C=1em @R=.7em {
& \qw & \multigate{1}{\phantom{U}} & \qw\\
& {/} \qw & \ghost{\phantom{U}} & \qw 
}   =\Qcircuit @C=1em @R=.7em {
& \qw & \qw & \gate{R_z} \qwx[1] & \qw & \gate{R_y} \qwx[1] & \qw & \gate{R_z} \qwx[1] & \qw & \qw\\
& {/} \qw & \gate{\phantom{U}} & \gate{} & \gate{\phantom{U}} & \gate{} & \gate{\phantom{U}} & \gate{} & \gate{\phantom{U}} & \qw
}   \nonumber
\end{align}
Regarding the three multiplexors, a two-qubit gate decomposition is known:
\begin{lem}[$k=x,y,z$,~\cite{shende2005synthesis,mottonen2004quantum,Bullock2004Asymptotically}]\label{lem:CCRNum}
The gate
\begin{align}
\Qcircuit @C=1em @R=.7em {
& \qw & \gate{R_k} \qwx[1] & \qw\\
& {/} \qw & \gate{} & \qw 
}\nonumber
\end{align}
can be implemented using $2^{n-1}$ two-qubit gates and single-qubit gates.
\end{lem} 
Based on this decomposition strategy, the minimum number $c_n$ of two-qubit gates required for $n$-qubit circuit synthesis satisfies the following recurrence relation:
\begin{align}
c_n \leq 4c_{n-1}+3\times 2^{n-1}.\nonumber
\end{align}
By iteratively applying the above recursive relation, we can compute $c_n$ as
\begin{align}
c_n \leq 4^{n-3}(c_3+3\times 2^2)-3\times 2^{n-1}.\nonumber
\end{align}
According to~\Cref{thm:3QubitRec}, we have $c_3\leq 11$. We can thus obtain $c_n\leq\frac{23}{64}4^n-\frac{3}{2}2^n$, which completes the proof of~\Cref{thm:CircSynRec}. Note that the ad hoc optimizations in Ref.~\cite{shende2005synthesis} are already exploited when we prove~\Cref{thm:3QubitRec}.

\end{document}